\documentclass[aps,prb,groupedaddress,showpacs,twocolumn]{revtex4}
\usepackage{graphicx}
\usepackage{amsmath}
\usepackage{amssymb}
\usepackage{dsfont}
\usepackage{bbm}
\usepackage{xspace}
\usepackage{color}
\usepackage{footnote}
\usepackage{comment}
\usepackage[caption=false]{subfig}
\usepackage{bibentry}
\usepackage[T1]{fontenc}

\newcommand{\avg}[1]{\langle #1\rangle}

\newcommand{\EE}{{\cal E}}

\newcommand{\abs}[1]{|#1|}
\newcommand{\tbd}[1]{}
\newcommand{\OLD}[1]{}
\newcommand{\NEW}[1]{}

\newcommand{\jncomm}[1]{}
\newcommand{\ea}[1]{}
\newcommand{\eacomm}[1]{}
\newcommand{\wvl}[1]{{\color{red} }}
\newcommand{\wvlcomm}[1]{}

\newcommand{\macomm}[1]{}

\newcommand{\fig}[1]{Fig.\thinspace{}\ref{#1}}

\newcommand{\Fig}[1]{Fig.\thinspace{}\ref{#1}}
\newcommand{\eq}[1]{eq.\thinspace{}(\ref{#1})}

\newcommand{\se}{Sec.\@\xspace}

\newcommand{\app}[1]{App.\@\xspace \ref{#1}}

\newcommand{\beq}{\begin{equation}}
\newcommand{\eeq}{\end{equation}}

\def\bra#1{\mathinner{\langle{#1}|}}
\def\ket#1{\mathinner{|{#1}\rangle}}

\newcommand{\nag}{{\phantom{\dag}}}

\begin{document}

\title{Current characteristics of a one-dimensional Hubbard chain: 
The role of correlation and dissipation}


\author{Jakob Neumayer}
\email[]{jakob.neumayer@tugraz.at}
\affiliation{Institute of Theoretical and Computational Physics, Graz University of Technology, NAWI Graz, 8010 Graz, Austria}
\author{Enrico Arrigoni}
\affiliation{Institute of Theoretical and Computational Physics, Graz University of Technology, NAWI Graz, 8010 Graz, Austria}
\author{Markus Aichhorn}
\affiliation{Institute of Theoretical and Computational Physics, Graz University of Technology, NAWI Graz, 8010 Graz, Austria}
\author{Wolfgang von der Linden}
\affiliation{Institute of Theoretical and Computational Physics, Graz University of Technology, NAWI Graz, 8010 Graz, Austria}


\date{\today}

\begin{abstract}
We study the electronic transport in an infinite one-dimensional Hubbard  chain, 
driven by a homogeneous electric field. The physical chain is coupled to fermion bath 
chains, in order to account for dissipation and to prevent the occurrence of Bloch Oscillations. 
The steady state current is computed in the frame of Keldysh Green's functions 
in Cluster Perturbation Theory. The current characteristics are dominated by resonant-tunneling-like 
structures, which can be traced back to Wannier-Stark resonances due to 
anti-ferromagnetic correlations.
The same current characteristic occurs in a non-interacting Wannier-Stark model with
alternating on-site energies. Non-local effects of the self energy can be accounted for the observed physical behaviour.     
\end{abstract}

\pacs{71.15.-m, 71.27+a, 71.10. -w, 73.23.-b}

\maketitle

\section{Introduction}\label{sec:introduction}

One dimensional (1d) quantum systems exhibit fascinating non-equilibrium effects such as
Bloch oscillations~\cite{f.bl.1928} (BOs), Zener tunneling~\cite{c.ze.1934}
and Wannier-Stark resonances.\cite{g.wa.1960, emin_phonon-assisted_1987, gluck_calculation_1998} The first theoretical studies of these effects 
date back to the early days of condensed matter physics and were followed by long-lasting controversial
discussions about their relevance in real materials.\cite{kr.ia.1986, nenciu.1991, bo.lu.1995} It was only within the last 
two decades that the theoretical predictions could be observed in condensed matter sytems and optical potentials 
in a series of ingenious experiments.\cite{fe.le.1992,ch.wa.1993, ya.mo.1994, bd.pe.1996,
wi.bh.1996, an.ka.1996, mo.mu.2001, he.ha.2001, mo.im.2002,me.ba.1993, ta.gr.2012}. 
A detailed overview over the most important experimental verifications of
the above mentioned effects is provided in Ref.~\onlinecite{gl.ko.02} and Ref.~\onlinecite{ha.mo.04}, where also experimental 
set-ups for determining Wannier-Stark resonances with cold atoms in optical lattices are introduced. 

Bloch oscillations are mainly studied in 1d systems which are and were frequently used as model systems to investigate the 
physical nature
of BOs. As far as correlations are concerned the 1d Hubbard model was mostly the model of choice.
The clear evidence for the existence of BOs and Wannier-Stark resonances led to a series of further theoretical 
investigations of 1d structures under the impact of a uniform force induced by an electric or magnetic 
field.\cite{gl.ko.02, ok.ar.03, wi.we.2004, arra.2004, ca.vo.94, mo.im.2002}

For the emergence of Wannier-Stark resonances in 1d systems a periodic potential has to be applied.
It was long thought that this periodic potential has to meet some
specific conditions.\cite{gr.ma.1993, gr.ma.1994,
gr.sa.1995, gr.sa.1997, gr.sa2.1997, gr.sa.1998, bu.gr.1998} Recently, it was shown
analytically that  
any periodic potential suffices to cause these resonant 
quantum states.\cite{sacchetti.2013}

The theoretical investigation of non-equilibrium properties have become very popular in recent 
years stimulated by the experimental progress and based on the development of powerful 
numerical methods that allow to tackle these challenging problems.
The nonequilibrium characteristics of 1d systems in a homogeneous electrical field, resulting in a linearly decreasing 
potential,
have been studied with various theoretical methods and in different geometries and gauges. One route of research considers 1d rings in the 
so-called temporal gauge, i.e. the ring is threaded by a magnetic flux that increases linearly in 
time. The advantage of this geometry is that one can use periodic boundary conditions.
For non-interacting particles the problem can be solved analytically
 for arbitrary ring size in the Keldysh formalism. For a review see Ref.~\onlinecite{arra.2004}.
It has been found that there is no dc-current possible without additional dissipation channels.
The latter can be introduced by attaching an infinite fermionic bath chain to each site of the ring.
As soon as electron-electron interactions are included, the problem can only be solved 
by numerical means and for very small ring sizes without dissipative bath chains. The real-time evolution for a 10-site ring 
has been studied~\cite{ok.ar.03} for  Hubbard interaction by the Cranck-Nicholson method and in Ref.~\onlinecite{es.co.2014} for the extended Hubbard model by the Chebyshev propagation method.
In both cases no  steady state dc current has been found, due to the lack of dissipation
processes. In Ref.~\onlinecite{es.co.2014}  it was shown that the current characteristics are always 
dominated by BOs.

Another class of studies considers a finite 1d Hubbard chain (central region) with linearly decreasing on-site energies, which 
are  attached on both ends to leads, represented by finite tight-binding (TB) chains of non-interacting electrons.
Time-dependent density-matrix renormalization group (tDMRG) calculations have been performed for a central region of 10 sites 
attached to 20 lead sites on both ends.\cite{he.go.10} The results are in qualitative agreement with  those found for 
the closed ring geometry,\cite{ok.ar.03} with the quintessence of an universal dielectric breakdown characteristic of the Mott 
insulating state. 
It is worth mentioning that a similar result was found by dynamical mean-field theory (DMFT) calculations on the 
hyper-cubic lattice.\cite{ec.ok.10}
However, in all real-time evolutions the current did not reach the steady-state.

Yet another possible setup to study the response of one dimensional lattice fermions 
to a homogeneous electric field is an infinite 1d TB-chain with linearly decreasing on-site energies with or without 
electron-electron interactions. In the non-interacting case the temporal gauge has been studied in great 
detail in Ref.~\onlinecite{han.13}, where a steady state was obtained 
by attaching fermionic dissipative baths to each site of the chain. The dissipation 
carries away the extra energy accumulated
and is essential to suppress BOs and to obtain a 
finite steady state current. 
Also in the two-dimensional one-band Hubbard model\cite{am.we.2012} an explicit coupling to 
fermion bath chains was
concluded to be essential for obtaining a description of non-equilibrium steady-states. It should be noted 
that in set-ups where a finite physical device is  attached to infinite leads, dissipation is already 
provided by the leads.\cite{im.la.1999,knap_nonequilibrium_2011,nuss_steady-state_2012}

Clearly, fermionic heat baths appear rather artificial at first glance.
However, a closer analysis reveals that their  impact
on a physical system is qualitatively the same as that of  
phonons.\cite{han.13}
Recent quantum many-body studies on this topic showed that many physical properties 
in model systems with fermion baths behave as predicted by the classical
Boltzmann transport theory.\cite{ar.ko.12, ha.li.13u, han.13}
Further arguments in favour of fermionic baths can be found in Refs.~\onlinecite{ha.li.13u}
or \onlinecite{ts.ok.09}, where the non-equilibrium steady-state of photo-excited correlated electrons has been studied.
The fermion bath model was also used 
earlier to describe the electron transport in a metallic ring,\cite{roy.2008} where it was 
denoted as an extended Büttiker's model,
It has to be mentioned here, that in some
systems dissipation mechanisms do not have to be included explicitly since they are included anyways by means of the setup.

In the present work, we will study an infinite Hubbard-chain in a homogeneous electric field.
The dissipation will be described  by fermionic baths\cite{han.13,am.we.2012,kotliar.han.2014}
attached to each site of the physical chain. This approach is particularly convenient from the theoretical point of view.
In contrast to previous works the non-equilibrium steady state of the interacting many-body system is
treated by a generalization of Cluster Perturbation Theory (CPT)\cite{ov.sa.89, se.pe.02}
to the non-equilibrium situation.\cite{knap_nonequilibrium_2011,nuss_steady-state_2012}
This allows to study the impact of dissipation and electron-electron interactions on 
the transport of lattice fermions in an infinite homogeneous electric field. Furthermore, non-local effects of the self 
energy can be included via CPT. It will be shown later, that the inclusion of those non-local effects is crucial to 
explain the resonant structures occurring in the current characteristics of the correlated Hubbard chain.

The paper is organized as follows: In \se~\ref{sec:method} we give an introduction to the model and its solution. The current 
characteristics in the non-interacting case are briefly summarized in \se~\ref{ssec:NonInteracting}
to allow for a comprehensive understanding of the current characteristics expected in infinite 1d structures. The effect of on-site
Coulomb interactions is discussed subsequently in \se~\ref{ssec:Interacting}. To gain a comprehensible physical explanation
of the  results, we describe an alternative way of modelling a non-interacting
system with a similar non-equilibrium behaviour as the interacting one in \se~\ref{ssec:ResonantTunneling}. 
We will close our considerations by a simplified nonequilibrium Variational Cluster Approach (VCA) treatment in \se~\ref{ssec:VCA}. 
The final conclusion is presented in \se~\ref{sec:conclusion}.

\section{THEORETICAL TREATMENT}\label{sec:method}

The setup of the model is depicted in \fig{fig:model}. 
Blue squares represent the correlated physical sites of the 1D Hubbard chain.
Each physical site is coupled via hopping to fermionic bath chains.
The parameters $t, t_{b},v$, as depicted in the figure,
stand for the hopping along the physical TB chain, within the bath chains, and the hopping between the physical system and the heat bath chains. The uniform electric field is applied parallel to the physical chain.

\subsection{The Hubbard-Wannier-Stark model}\label{ssec:Model}
\begin{center}
 \begin{figure}
\includegraphics[width=0.48\textwidth]{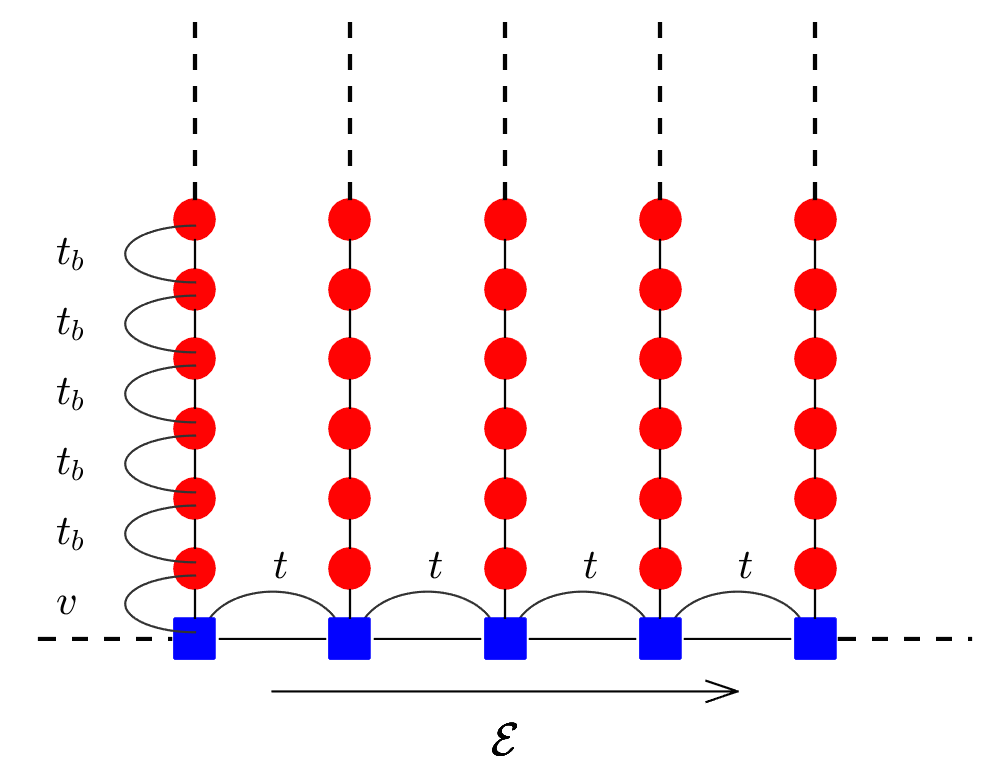}
\captionsetup{justification=raggedright,singlelinecheck=false}
\caption{(Color online) Sketch of the infinite one-dimensional correlated chain (blue squares), coupled at each site to individual 1d heat bath chains (red circles).
The homogeneous electric field 
${\cal E}$ is applied parallel to the physical chain. }
\label{fig:model}
\end{figure}
\end{center}
The Hamiltonian corresponding to the system depicted in \fig{fig:model} is reads
\begin{align}
\mathcal{H} &= \mathcal{H}_{\text{0}} + \mathcal{H}_{\text{U}} + \mathcal{H}_{\text{bath}}\label{eq:H}\;.
\end{align}
The kinetic part of the correlated chain is described by a  TB-Hamiltonian 
\begin{align}
 \label{eq:H_TB}
\mathcal{H}_{\text{0}} &= -t  \sum\limits_{\left<i,j\right>,\sigma}  {c}_{i\sigma}^{\dagger}  {c}_{j\sigma}^{\nag}\, + 
\sum\limits_{j} \epsilon_{j}  {n}_{j}  \, \mbox{,}
\end{align}
where, as usual, ${c}_{j\sigma}^\dag$ (${c}_{j\sigma}^\nag$) denote fermionic
creation (annihilation) operators,  for  site  $j$ and  spin  
$\sigma\in\{\uparrow,\downarrow\}$, and
${n}_{j}=\sum_{\sigma}{c}_{j\sigma}^\dag {c}_{j\sigma}^\nag$
stands for the particle number operator. 
In  this paper we use $t=1$.
In addition, we also set the lattice constant and the electronic charge to one.
The on-site energies $\epsilon_{i}$  contain the linear potential due to the applied homogeneous electric field ${\cal E}$

\begin{align}
\label{eq:epsilon}
\epsilon_{j} &= \bigg(j-\frac{1}{2}\bigg) \;\EE \;.
\end{align}

The electron-electron interaction present in the physical chain 
is modelled  by the Hubbard Hamiltonian
\begin{align}
\label{eq:H_IA}
\mathcal{H}_{\text{U}} &= \frac{U}{2} \, \sum\limits_{i\sigma} 
\big({n}_{i\overline{\sigma}}-\frac{1}{2} \big)\, \big({n}_{i\sigma} -\frac{1}{2}\big)
\end{align}
where $U$  represents the repulsive on-site Coulomb interaction. 

The dissipation that is mediated by  fermionic heat-bath chains is  described by 
individual TB chains attached to each physical site. The corresponding part of the Hamiltonian reads
\begin{align}\label{eq:H:bath}
\mathcal{H}_{\text{bath}} = \, & - t_b  \sum\limits_{\alpha j \sigma}  \, 
\left({d}_{\alpha j \sigma}^{\dagger} \, {d}_{\alpha+1, j\sigma}^{\nag}\, + h.c.\right)
+ \sum\limits_{\alpha j\sigma} \epsilon_{j} \, n^{d}_{\alpha j \sigma}     \\ \notag
&- v \, \sum\limits_{j \sigma} \, \left( {c}_{j\sigma}^{\dagger} \, {d}_{1 j\sigma}^{\nag}\, + h.c.\right) \;,
\end{align}
where ${d}_{\alpha j\sigma}^{\dag}$ and ${d}_{\alpha j\sigma}^{\nag}$ are  creation and annihilation operators for particles with spin $\sigma$ at position $\alpha$ within
the j$^{th}$ bath chain, respectively, and $n^{d}_{\alpha j\sigma}$ is the corresponding number operator. 
The third term represents the hybridization between the $j^{th}$ site in the physical chain and the corresponding bath chain. 
As we will see later it is only the ratio of the two external parameters $t_B$ and $v$ that affects our results. Therefore we
have used $v=t$ throughout this paper. Note that the thermal bath chains experience the same E-field potential energy $\epsilon_{j}$ as the corresponding physical site. Otherwise the hybridization would violate
the gauge invariance as discussed in \app{app:GaugeTransformation}. The bath chains are considered to be in an equilibrium state
infinitely far away from the correlated chain with different chemical potentials $\mu$ given by

\begin{align}
\label{eq:epsilon}
\mu_{j} &= \epsilon_j = \bigg(j-\frac{1}{2}\bigg) \;\EE \;.
\end{align}

In the Keldysh Green's function formalism the current $j_{mn}$ 
between two adjacent lattice sites $m$ and $n$ 
for the aforementioned Hamiltonian
can be obtained from the Keldysh component of the  non-equilibrium Green's function:
\begin{align}
\label{eq:CurrentDensity}
j_{mn} &= -\frac{t}{2\pi} \int_{-\infty}^{\infty} d\omega \text{ }  \Re\left\{ G_{nm}^K(\omega) \right\} \, \mbox{.}
\end{align}
In the next sections
we outline how the Keldysh Green's function for the infinite chain for interacting electrons
in a homogeneous electric field can be determined.

\subsection{Cluster Perturbation Theory\label{ssec:Model}}

One way to treat 
properties of 
interacting many-body systems is 
Cluster Perturbation Theory (CPT).\cite{ov.sa.89, se.pe.02}
For non-interacting particles, CPT  always gives the exact result. 
In the presence of electron-electron interaction CPT is no longer exact; 
it is a first order perturbation theory in the inter-cluster hopping parameters. 
The accuracy of CPT increases with the size of the clusters that are treated exactly by numerical means.

\begin{center}
 \begin{figure}[h!]
 
 \subfloat[]{
\includegraphics[width=0.148\textwidth]{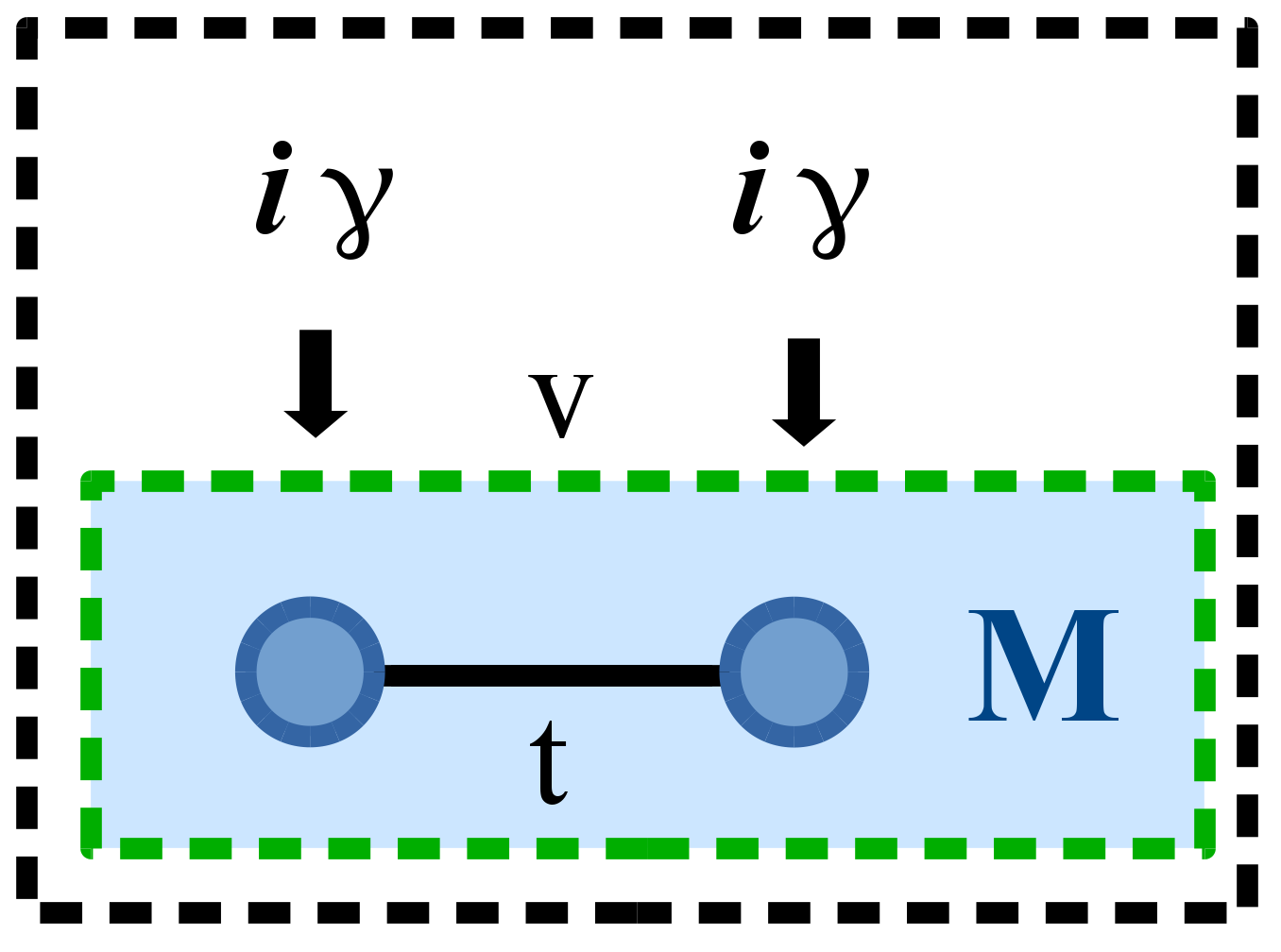}}  \hspace{0.035\textwidth}
 \subfloat[]{
\includegraphics[width=0.268\textwidth]{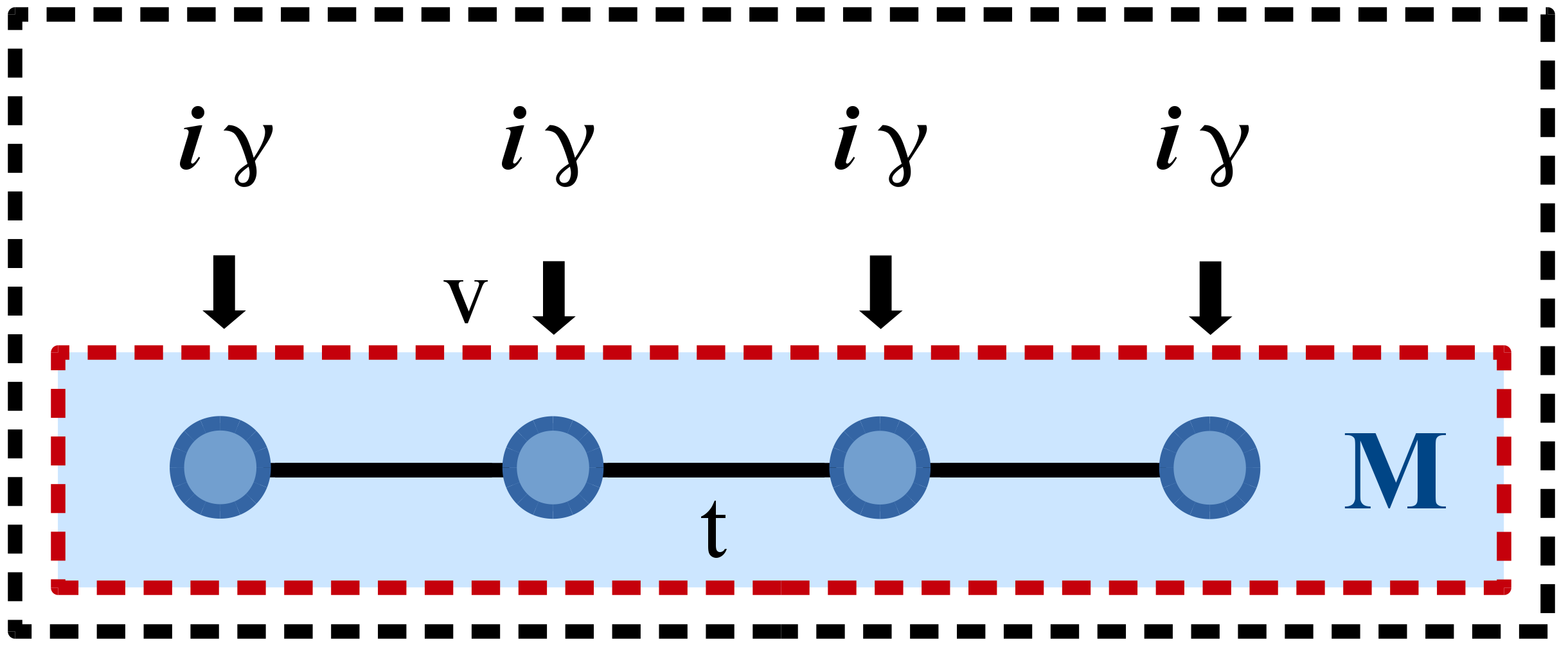}} \\
 \subfloat[]{
\includegraphics[width=0.4\textwidth, height=0.1\textwidth]{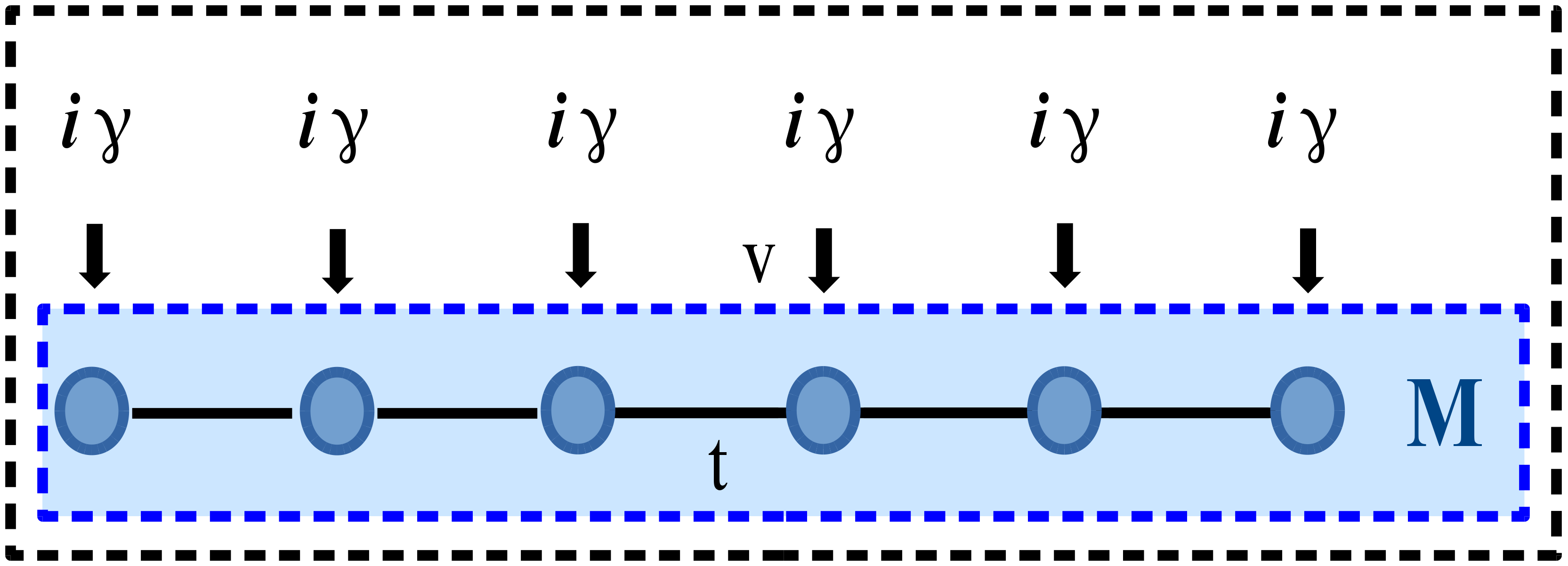}}  
\captionsetup{justification=raggedright,singlelinecheck=false}
\caption{(Color online) Central clusters of size $M=2$ (a), $M=4$ (b) and $M=6$ (c), which in CPT are
periodically strung together to form the infinite physical chain. The dissipation mechanism is represented by the wide-band
result $i \gamma$.}
\label{fig:model2}
\end{figure}
\end{center}

For the model at hand, which still has a kind of translational invariance, it suffices to perform an exact diagonalization 
for one 'central cluster'. The Green's function of the central cluster
is then glued together iteratively by CPT to form the infinite system.  Possible central clusters
are depicted in \Fig{fig:model2}.

\subsection{Non-equilibrium CPT for the Keldysh Green's functions}\label{ssec:KNEGF}

The CPT approach can be generalized to Keldysh Green's Functions.\cite{keld.65}
In previous studies\cite{knap_nonequilibrium_2011,nuss_steady-state_2012} a finite device with interacting electrons has been 
coupled to infinite leads. Here we will generalize these ideas to  infinite Hubbard-chains subjected to a homogeneous 
electric field. 

The single-particle Green's functions in this generalization are represented as four-component matrices in 
Keldysh space\cite{kad.baym,schw.61,keld.65,ha.ja,ra.sm.86} 
\begin{align}
\label{eq:KeldyshSpace}
\boldsymbol{g}(\omega) & = \begin{pmatrix} g^{r}(\omega) & g^{k}(\omega) \\ 0 & g^{a}(\omega) 
\end{pmatrix} \,\mbox{.}
\end{align}
Lower case $g$ denotes a Green's functions of an isolated cluster here, while 
the labels $r$, $a$ and $k$ stand for retarded, advanced and Keldysh component, 
respectively. Bold letters are used for the full Keldysh  Green's function given in \eq{eq:KeldyshSpace}. 
The size of the sub-matrices $g^{r}$, $g^{a}$ and $g^{k}$ is determined by the size of the 
chosen cluster, as we will see later. Note that this notation applies throughout the whole paper. 
We will neglect the explicit frequency-dependence of the Green's functions 
in the following for reasons of readability and will re-introduce it where necessary.

In equilibrium, the Keldysh component is related to the advanced/ retarded components via

\begin{align}
\label{eq:GKeldysh}
g^k &= \left[g^r - g^a\right] \left[2f_F(\omega, \mu, T)-1\right] \, \mbox{,}
\end{align}
where $f_F(\omega, \mu, T)$ denotes the Fermi function depending on the chemical potential $\mu$ and the temperature $T$.

\vspace{0.5cm}
\begin{center}
 \textbf{General scheme of CPT}
\end{center}

The basic step of the theory is the coupling of two initially decoupled isolated  parts, which we will
label by $\alpha,\beta\in\{1,2\}$. For this system the Green's function is 
cluster diagonal, i.e.
$\boldsymbol{g}^{\alpha\beta}=\delta_{\alpha\beta}\boldsymbol{g}^{\alpha\alpha}$.
Now the CPT approximation for the Keldysh Green's functions $\boldsymbol{G}$ 
of the whole coupled system, written in 
form of a Dyson equation,
reads 

\begin{subequations}\label{eq:CPT:Dyson:Keldysh0}
\begin{align}
\boldsymbol{G}^{\alpha\beta}  &= \delta_{\alpha\beta}\boldsymbol{g}^{\alpha\alpha} + \sum_{\overline{\alpha} \in \alpha,\beta}
\boldsymbol{g}^{\alpha\alpha} * \boldsymbol{T}^{\alpha\overline{\alpha}} * \boldsymbol{G}^
{\overline{\alpha}\beta}\;,\\
\boldsymbol{T}^{\alpha\overline{\alpha}} &= 
\begin{pmatrix}
  T^{\alpha\overline{\alpha}}&0\\
 0& T^{\alpha\overline{\alpha}}
\end{pmatrix}\;.
\end{align}
\end{subequations}

where the matrix multiplications have to performed in Keldysh space (see \eq{eq:KeldyshSpace}). 
$T^{\alpha\overline{\alpha}}$ stands for the matrix that contains the parameters for the hopping processes  
that connect the two clusters $\alpha$ and $\overline{\alpha}$.
In the 1d case with only two clusters considered here, the only non-zero matrix elements of $T^{\alpha\overline{\alpha}}$ are 
given by
\begin{equation*}
 T^{1,2}_{1,M} = -t = T^{2,1}_{M,1}\,
\end{equation*}
where $M$ denotes the size of the cluster.

We are interested in the sub-matrix of $\boldsymbol{G}$ belonging to the first cluster only, i.e. where $\alpha=\beta=1$.
Evaluating \eq{eq:CPT:Dyson:Keldysh0} leads to

\begin{subequations}\label{eq:CPT:Dyson:Keldysh}
\begin{align}
\boldsymbol{G}^{11}  &= \boldsymbol{g}^{11} + \boldsymbol{g}^{11}
\boldsymbol{\Gamma}^{11} \boldsymbol{G}^{11}\\
\boldsymbol{\Gamma}^{11} &:= \boldsymbol{T}^{12} \boldsymbol{g}^{22}\boldsymbol{T}^{21} 
\end{align}
\end{subequations}

A Dyson equation in Keldysh space of the form

\begin{align}\label{eq:Dyson:symbolic}
\boldsymbol{G}  &= \boldsymbol{g} + \boldsymbol{g}\,\boldsymbol{\Gamma} \,\boldsymbol{G}\;
\end{align}

can be solved for the advanced/ retarded and Keldysh component separately.
For the retarded component  we find 

\begin{align}\label{eq:CPT:Dyson:ret}
\big(G^{r}\big)^{-1} &= \big(g^{r}\big)^{-1} - \Gamma^{r}\;,
\end{align}

which has the same form as in equilibrium. Advanced and retarded Green's function are
related through $G^{a}=\big(G^{r}\big)^{\dagger}$.
For the Keldysh component we obtain a special form of the Kadanoff-Baym equation\cite{ha.ja}
\begin{align}\label{eq:CPT:Dyson:Keldysh:KKB}
G^{k} &= 
G^{r}
\left\{
\big(g^{r}\big)^{-1}
g^{k}
\big(g^{a}\big)^{-1}
+
 \Gamma^{k}
\right\}
G^{a}\;.
\end{align}

If $ g$ belongs to a finite size cluster in equilibrium then the first term in curly brackets
can be neglected.
Inserting \eq{eq:GKeldysh} into \eq{eq:CPT:Dyson:Keldysh:KKB} yields $\big(g^{a}\big)^{-1}-\big(g^{r}\big)^{-1}=2i0^+$.
In the applications in the present paper the second term will
originate from semi-infinite TB-chains in the wide-band limit. It will therefore contribute a finite value for 
$\Gamma^{k}$ much larger than $0^+$, so that the first term can be ignored.

\vspace{0.5cm}
\begin{center}
 \textbf{Coupling to dissipative chains}
\end{center}

Now we begin with the construction of the Keldysh Green's function for the infinite system 
in an electric field.
We start out  with the computation of the Keldysh Green's function $\boldsymbol{g^{cl}}(\omega)$  
for the $M$-site central cluster based on Exact Diagonalization. 
In this approximation the initially uncoupled parts are in equilibrium. Therefore it suffices to compute the retarded 
Green's function.

We now use CPT to couple the central cluster to bath chains which are attached according to \fig{fig:model}.
We are only interested in Green's functions $\boldsymbol{G}^{cc}$, where both indices $\alpha$, $\beta$ in 
\eq{eq:CPT:Dyson:Keldysh0} belong to the central cluster. 
In this case \eq{eq:CPT:Dyson:Keldysh} reads
\begin{subequations}\label{eq:CPT:Dyson}
\begin{align}
\boldsymbol{G}^{cc} &= \boldsymbol{g}^{cc} + \boldsymbol{g}^{cc}  
\boldsymbol{\Gamma} \boldsymbol{G}^{cc}\\
\boldsymbol{\Gamma} &:= T^{cb} \boldsymbol{g}^{bb} T^{bc}\;,
\end{align}
\end{subequations}
where $b$ denotes the bath chains, see Fig.~1.
The  matrix $\boldsymbol{\Gamma}$ describes the dissipation and de-phasing of the transport along the physical chain.
Due to the nearest neighbour hopping along the bath chains 
the sub-matrices $g^{bb,\zeta}$ of $\boldsymbol{g}^{bb}$ , where $\zeta$ stands for retarded, advanced or Keldysh,  are diagonal, i.e
\begin{align}
g^{bb.\zeta} &= \text{diag} \big(g^{b.\zeta}_{1},g^{b.\zeta}_{2},\ldots,g^{b.\zeta}_{M}\big)
\end{align}
with
\begin{align}
g^{b.\zeta}_{j}(\omega) &=g^{b.\zeta}(\omega- \epsilon_{j},\mu_{j}).
\end{align}
Here $g^{b.\zeta}(\omega, \mu_j)$
is the local Green's function of the semi-infinite bath that is connected to site $j$ of the central cluster at the first contact point. $\epsilon_{j}$ is the on-site energy and $\mu_{j}$ the 
chemical potential in the $j$-th bath chain. At half filling we have $\mu_{j}= \epsilon_{j}$. The chemical potential only 
enters the Keldysh component. For the retarded part 
of the semi-infinite TB Green's function we have\cite{economou.06} 
\begin{align}
\label{eq:GBath}
g^{b.r}(\omega) =& \frac{1}{t_B} \left\{ \frac{\omega}{2 t_b} - i\sqrt{1-\left(\frac{\omega}{2 t_b}\right)^2}\right\} 
\, \mbox{,}
\end{align}
with
\begin{align*}
\Im\left\{g^{b.r}(\omega)\right\} \equiv  0 \quad \text{for} \quad |\omega| \geq 2 t_B
\, \mbox{.}
\end{align*}

The Keldysh component $g^{b.K}$ follows from \eq{eq:GKeldysh} as
\begin{align}
g^{b.k}(\mu_j) &= \left[g^{b.r} - \left(g^{b.r}\right)^\dagger\right] 
\left[2f_F( \mu_j, T)-1\right] \, \mbox{.}
\end{align}

In the wide-band limit\cite{han.13, ha.li.13u, ar.ko.12}  $t_{b} \gg \abs{\omega} $ we find
\begin{align*}
g^{b,r}(\omega) =& -\frac{i}{t_{b}}\;.
\end{align*}
In this case, the self-energy $\boldsymbol{\Gamma}$ entering the Dyson
equation (\eq{eq:CPT:Dyson}) 
becomes
\begin{align}\label{eq:GammaEff}
\Gamma=iv^{2}/\abs{t_{b}}:=i\gamma\;.
\end{align}
That means that the dissipation is be described by a constant dissipation parameter
$\gamma$.
More precisely, the retarded part of the cluster Green's function follows from that of the isolated cluster by
\begin{align}\label{eq:CPT:Dyson:ret}
\big(\tilde{G}^{cc.r}\big)^{-1} &= \big(\tilde{g}^{cc.r}\big)^{-1} + i \gamma\;.
\end{align}

\vspace{0.5cm}
\begin{center}
 \textbf{Construction of the infinite system}
\end{center}

\begin{figure}[b]
 \subfloat{
\includegraphics[width=0.3\textwidth]{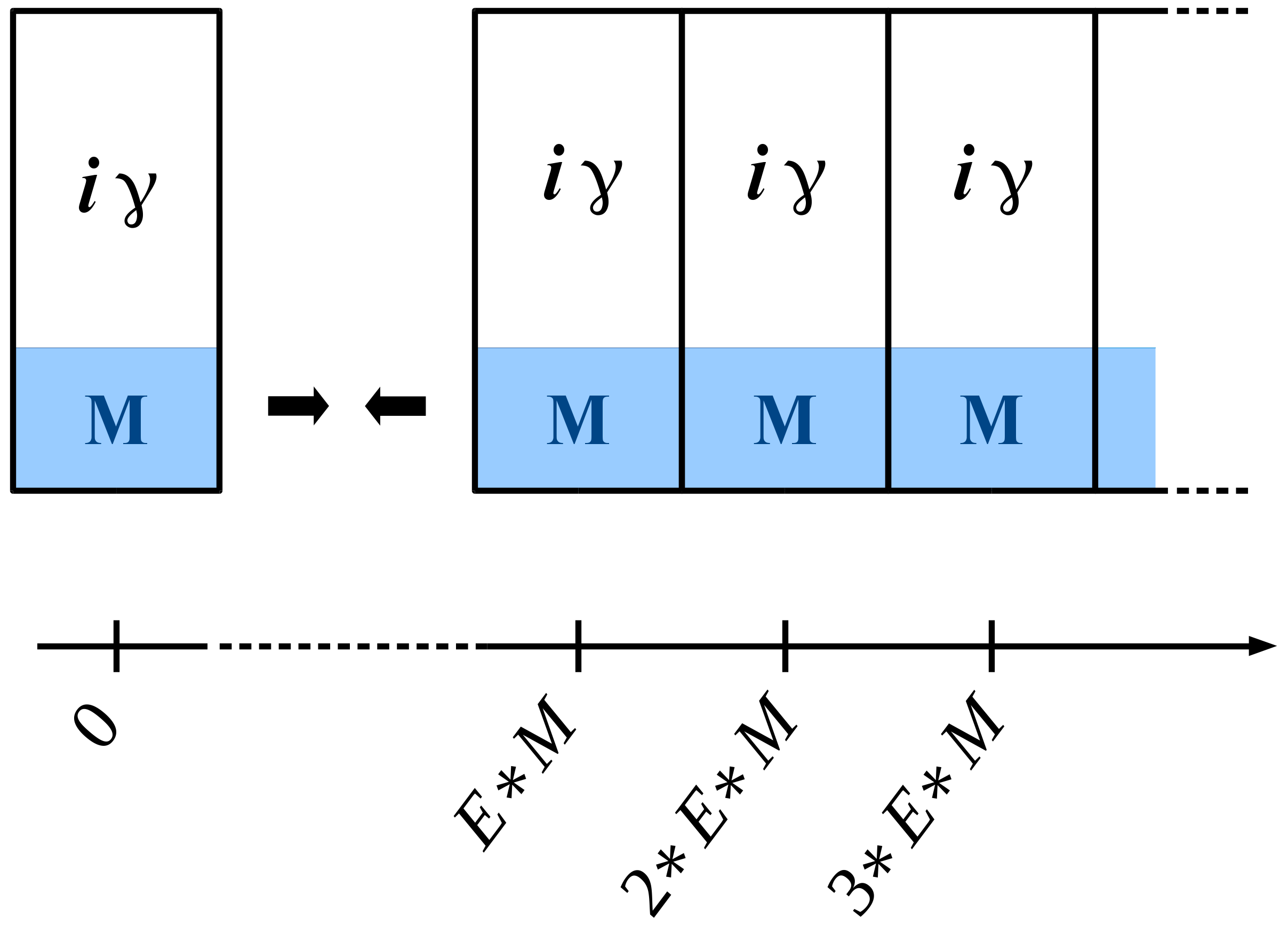}} \\
\caption{(Color online) Iterative construction of the right semi-infinite system.  
The values of the potential energy $\epsilon_{i}$ due to the applied electric field are depicted below (lattice constant 
and electronic charge are set to unity).}
\label{fig:Halfinfinite}
\end{figure}

Given the Green's function of the central cluster we can iteratively combine these clusters to form the infinite chain.
This can be done best by first determining a right/left  semi-infinite 'chain' of these clusters with the coupled bath chains 
and finally connecting them 
to end up with the desired infinite structure.

We begin with the construction of the semi-infinite part on the right half of the system. Let
 $\boldsymbol{G}^{11}(\omega)$ stand for the corresponding Green's function in which the site indices are restricted 
 to the first (leftmost) cluster of the semi-infinite part. 

The best way to obtain an iteration equation for this Green's function is given by
connecting a dissipative central cluster - with its Green's function $\boldsymbol{G}^{cc}$ - to the first cluster of the 
remaining semi-infinite part by CPT (see \fig{fig:Halfinfinite}).
The coupled system then is the same as the original one 
apart form an overall energy shift \hbox{$\omega_{s} = \EE M$}. Therefore, \eq{eq:CPT:Dyson:Keldysh} yields

\begin{align}
\boldsymbol{G}^{11}(\omega)   &=  \boldsymbol{G}^{cc}(\omega) 
\left[1+  
\boldsymbol{T}^{12}\boldsymbol{G}^{11}
\left(\omega  \!- \!\omega_{s} \right)\boldsymbol{T}^{21}\,
\boldsymbol{G}^{11}(\omega)\right]\;.
\end{align}

\begin{figure}[h]
 \subfloat{
\includegraphics[width=0.5\textwidth]{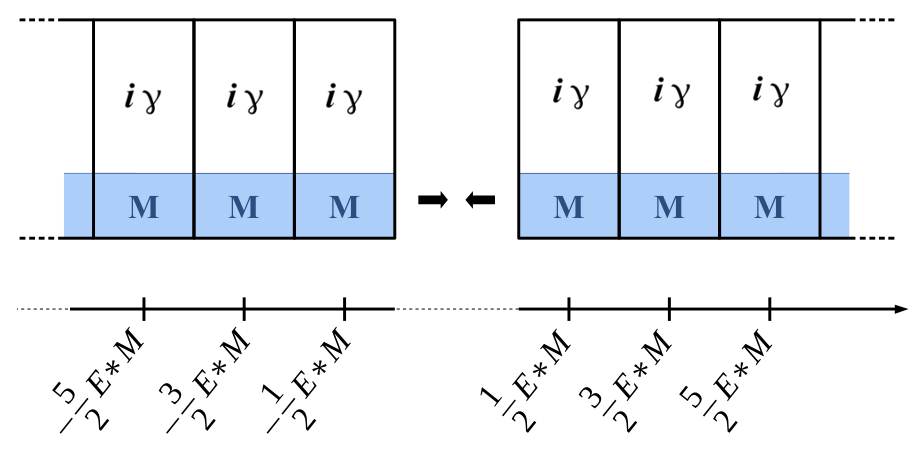}} \\
\caption{(Color online) Decomposition of the infinite system into two semi-infinite parts. 
The zero point of the potential energy is chosen in the middle of the 
central part. }
\label{fig:Decomposition}
\end{figure}

This equation can be solved by considering it as an iteration equation starting
with $\boldsymbol{G}^{11}(\omega) = \boldsymbol{G}^{cc}(\omega)$.
We obtain a similar equation for the left semi-infinite chain with 
$\omega_s= - M \EE$. Finally the two semi-infinite chains (clusters) can be combined again by 
CPT  based on \eq{eq:CPT:Dyson:Keldysh} (see \fig{fig:Decomposition}).

\section{Results}\label{ssec:Results}

\subsection{Non-interacting case}\label{ssec:NonInteracting}

Now we turn to the results obtained with the Keldysh CPT 
for the model employed in the present paper. First we present a short summary of
the key results in the non-interacting case and compare the description of the dissipative bath chains by semi-infite TB chains
to the wide band limit approach. 

An instructive approximation is obtained for
$\abs{\EE},\gamma\ll \abs{t}$ and wide-band limit, where  one finds\cite{han.13} 

\begin{align}\label{eq:j:han}
j(\EE)&\approx \frac{2 t a q}{\pi} \frac{\frac{\EE}{2\gamma}}{1+\big(\frac{\EE}{2 \gamma}\big)^{2}}\;.
\end{align}

Here $q$ is the charge of the particles and $a$ the lattice constant.It should be re-emphasized that we use units in 
which $q=1,a=1,t=1$.
This approximate result has a universal  maximum, characterized by 
\begin{subequations}{\label{eq:j:han:max}}
\begin{align}
\EE_\text{max} &= 2 q a\Gamma\;,\\
j_\text{max} &= \frac{t}{\pi}\;.
\end{align}
\end{subequations}

As emphasized before, in the non-interacting case CPT provides the exact result, irrespective of the 
cluster size. Therefore we here also
compare our results obtained by the Keldysh CPT approach to previous works.\cite{han.13, ar.ko.12, ha.li.13u, ts.ok.09}
And indeed for $U = 0$ we find that our CPT results 
for the infinite non-interacting TB-chain are 
in perfect agreement  with those for the same system obtained in temporal gauge.\cite{han.13}
For non-interacting particles and in the wide-band limit of the bath-chains it turns out that the coupling of the fermionic 
bath-chains modifies 
the cluster (advanced) self-energy by a constant damping $\gamma$, as already indicated by \eq{eq:CPT:Dyson:ret}. 

\begin{center}
 \begin{figure}[t!]
\includegraphics[width=0.4\textwidth]{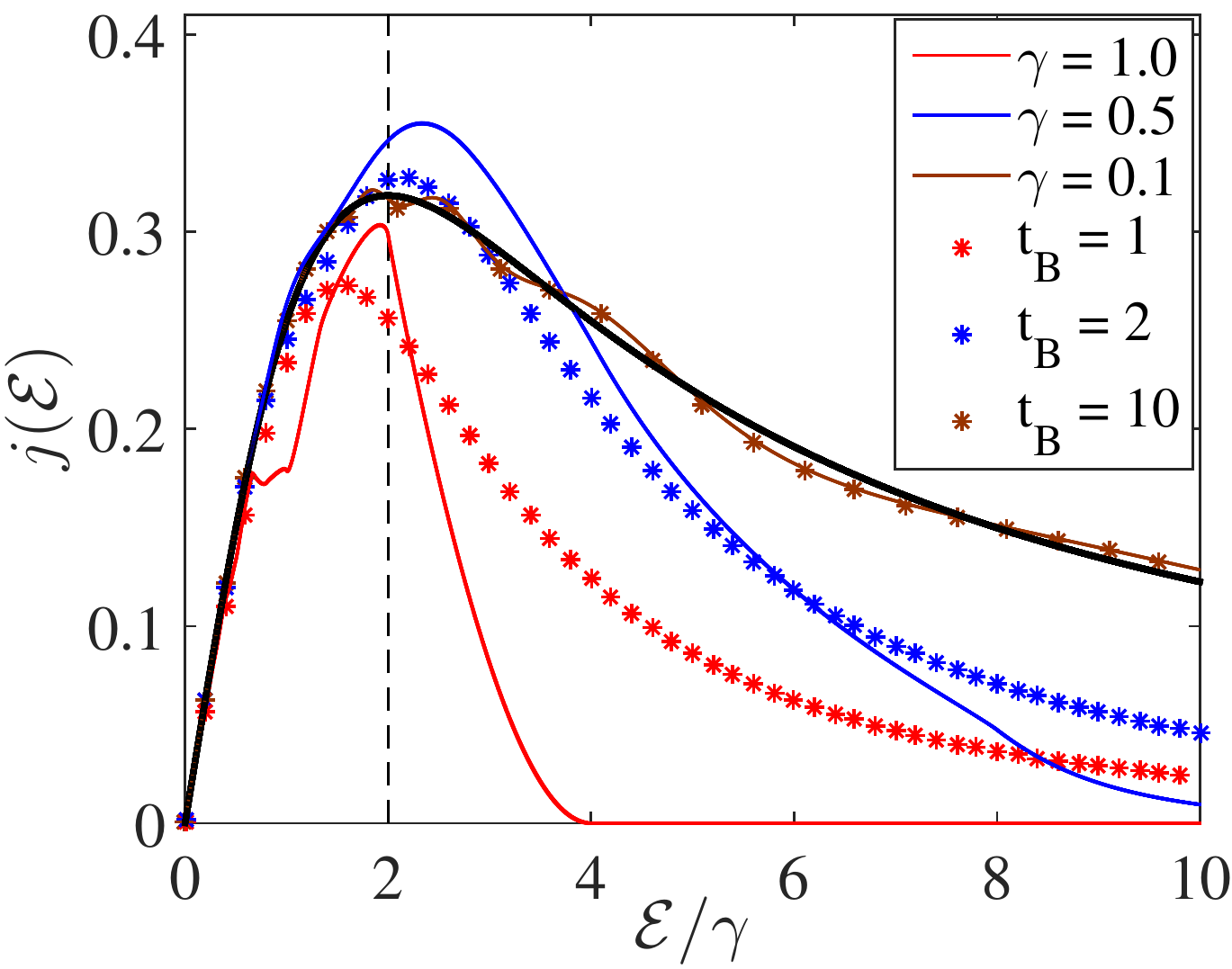}
\captionsetup{justification=raggedright,singlelinecheck=false}
\caption{(Color online) Current  calculated according to \eq{eq:CurrentDensity} in the non-interacting case ($U = 0$). Shown
with full lines are the results for dissipation described by a wide-band limit approach. In comparison the results for dissipation
described by semi-infinite TB chains are shown with star markers. The approximate current gained by evaluating \eq{eq:j:han}
is plotted for comparison (black line). The position of its maximum is marked by a dashed line.}

\label{fig:CurrentNonInteracting}
\end{figure}
\end{center}

\Fig{fig:CurrentNonInteracting} compares the current $j(\EE)$ in the physical chain
as a function of the electric field $\EE$ for the two different approaches for the dissipative bath chains. Curves labeled by
the value for the constant damping $\gamma$ correspond to the wide-band limit approach, while the ones where dissipation is 
described by semi-infinite TB chains are labeled by the value of $t_B$. 
The behaviour of the current is characterized by a pronounced uni-model structure.
While peak position and height are reasonably well described for all sets of parameters by the approximation
in \eq{eq:j:han:max}, one 
observes  significant deviations in the current curve for large values of $\gamma$ and small values of $t_B$ respectively. 
However, already for $\gamma=0.1$ ($t_B = 10$) we find an almost  perfect quantitative agreement even up to $\EE=1$, 
where the condition $\EE\ll t$ is not really met. Since we will restrict our further investigations to the $\gamma<0.1$ regime,
the wide-band limit approach chosen in the following is completely justified.

 \begin{figure*}[t!]
\includegraphics[width=0.32\textwidth]{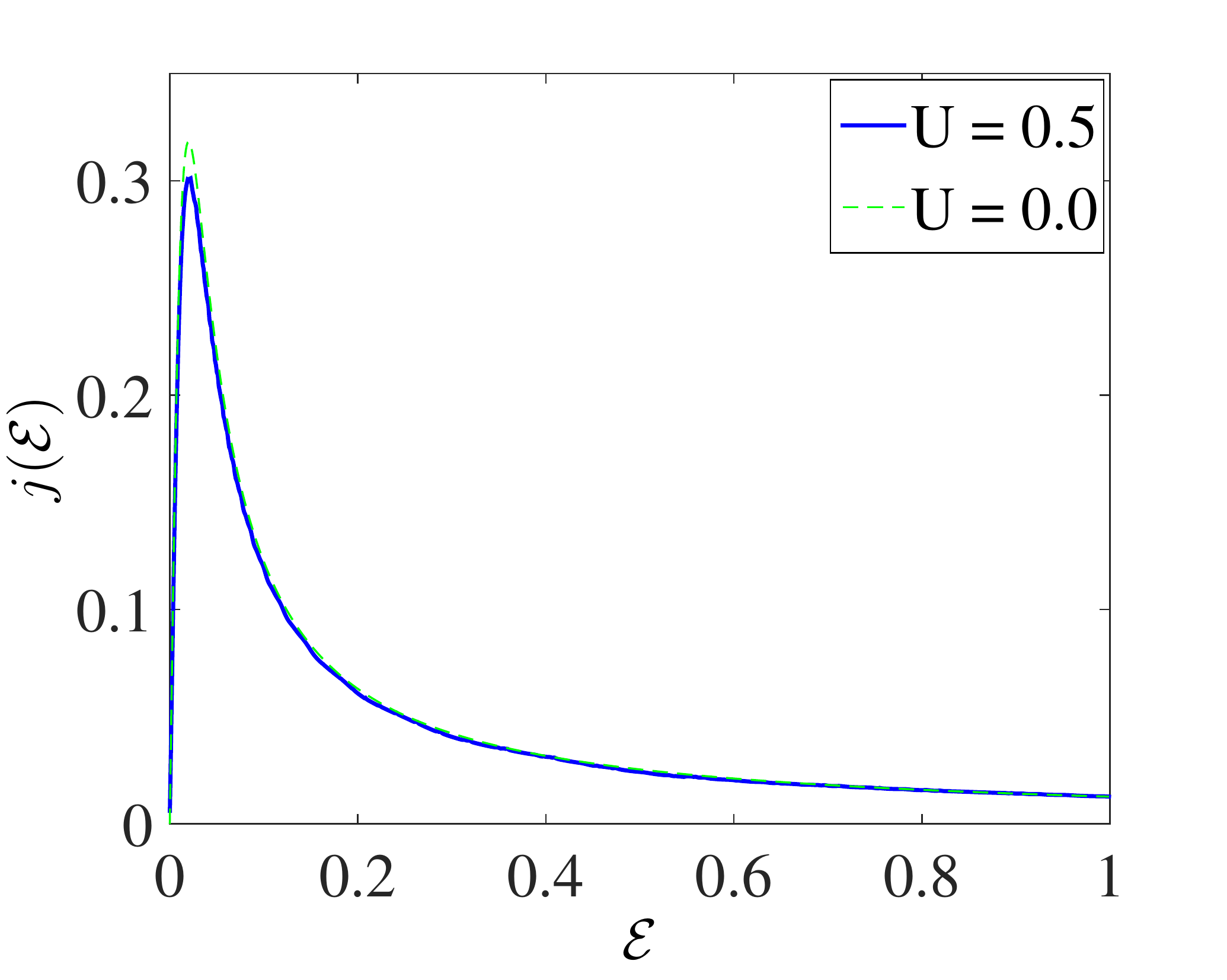}
\includegraphics[width=0.32\textwidth]{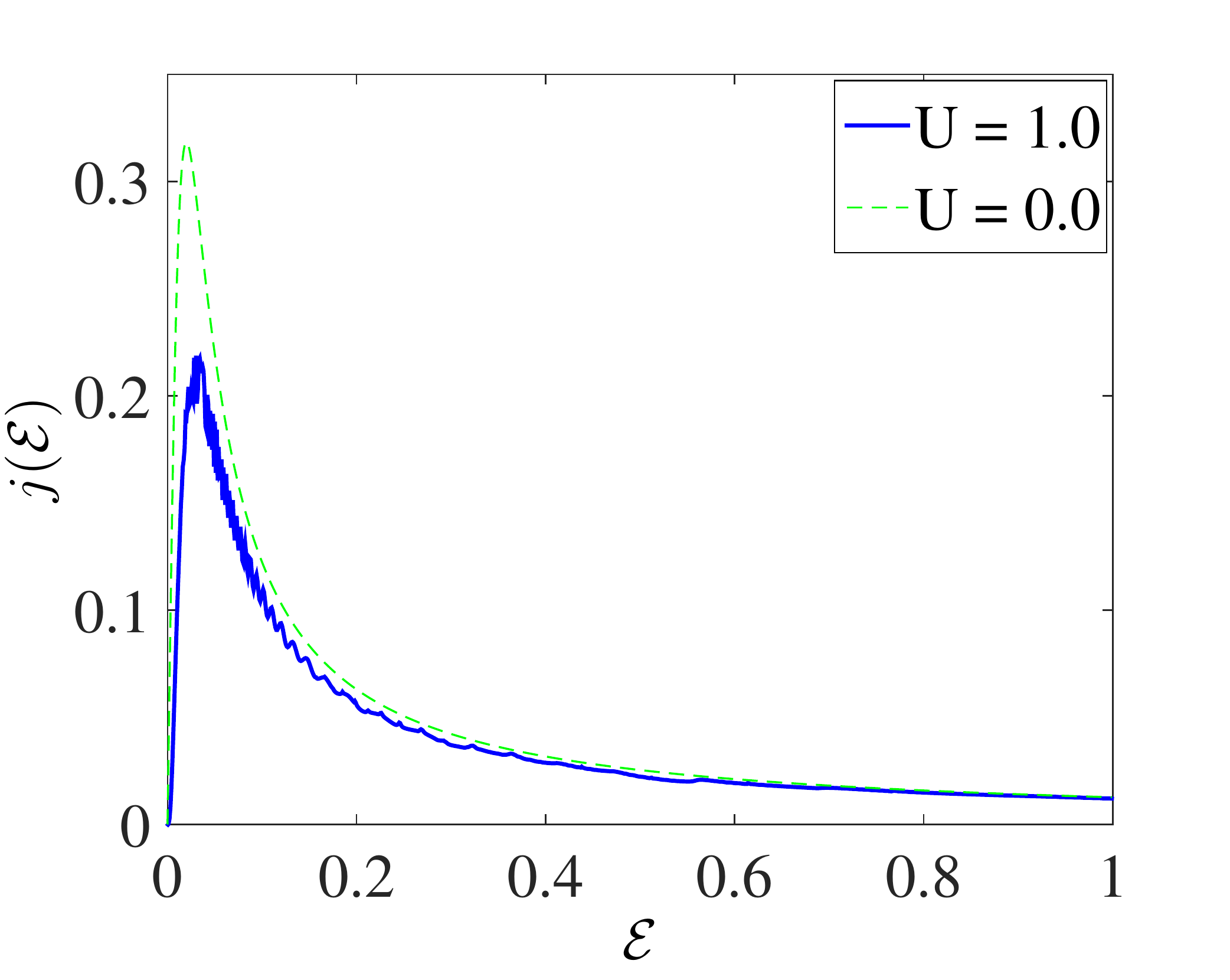}
\includegraphics[width=0.32\textwidth]{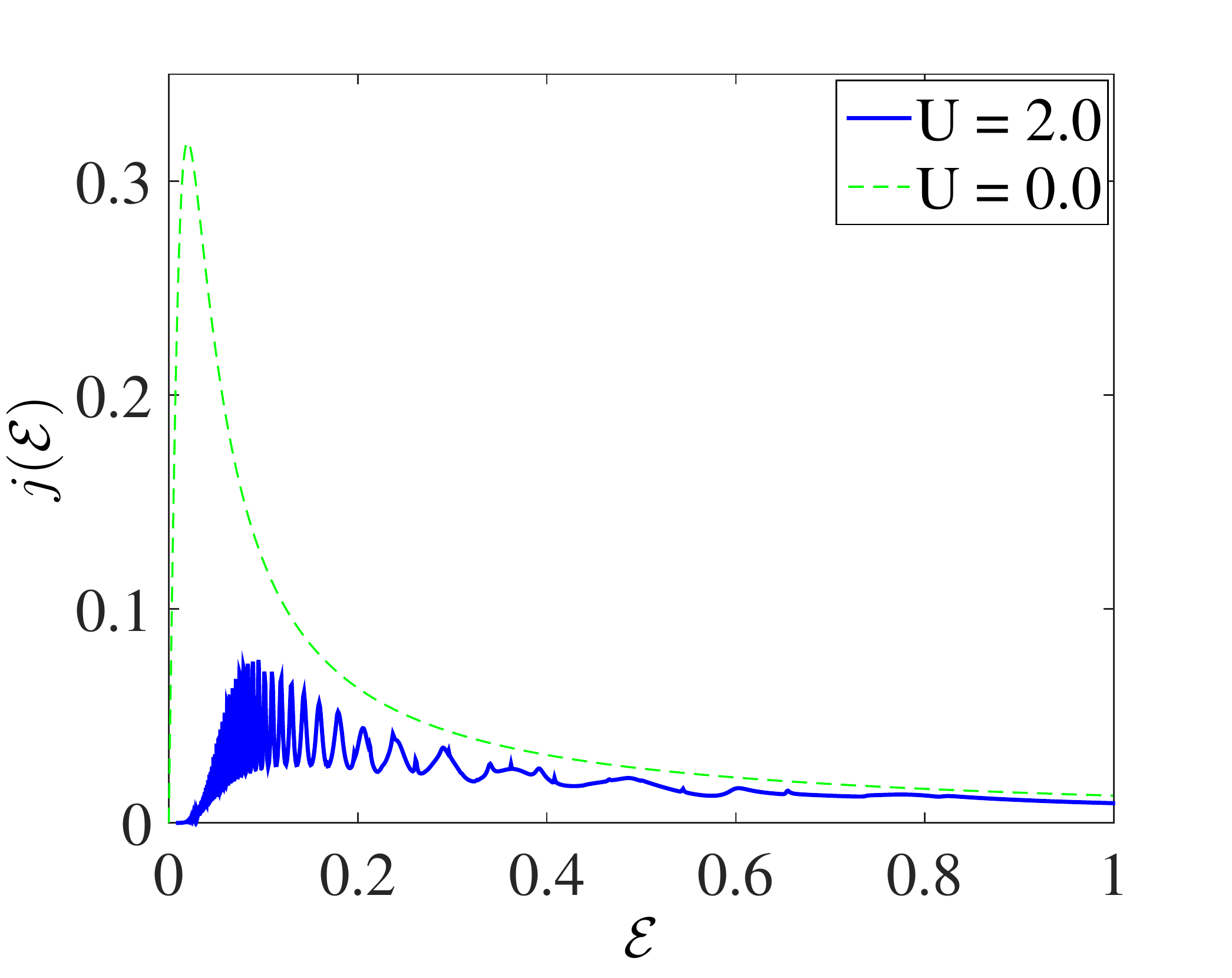}

\captionsetup{justification=raggedright,singlelinecheck=false}
\caption{(Color online) Current $j(\EE)$ (solid line, calculated according to \eq{eq:CurrentDensity}) for different values of 
the interaction strength $U$ = 0.5, 1.0, 2.0 and an effective damping $\gamma = 
0.01$.
The $U=0$ result is shown as reference in every subplot (green dashed line).}
\label{fig:CurrentInteracting}
\end{figure*}
 \begin{figure*}[t!]
 \includegraphics[width=0.32\textwidth]{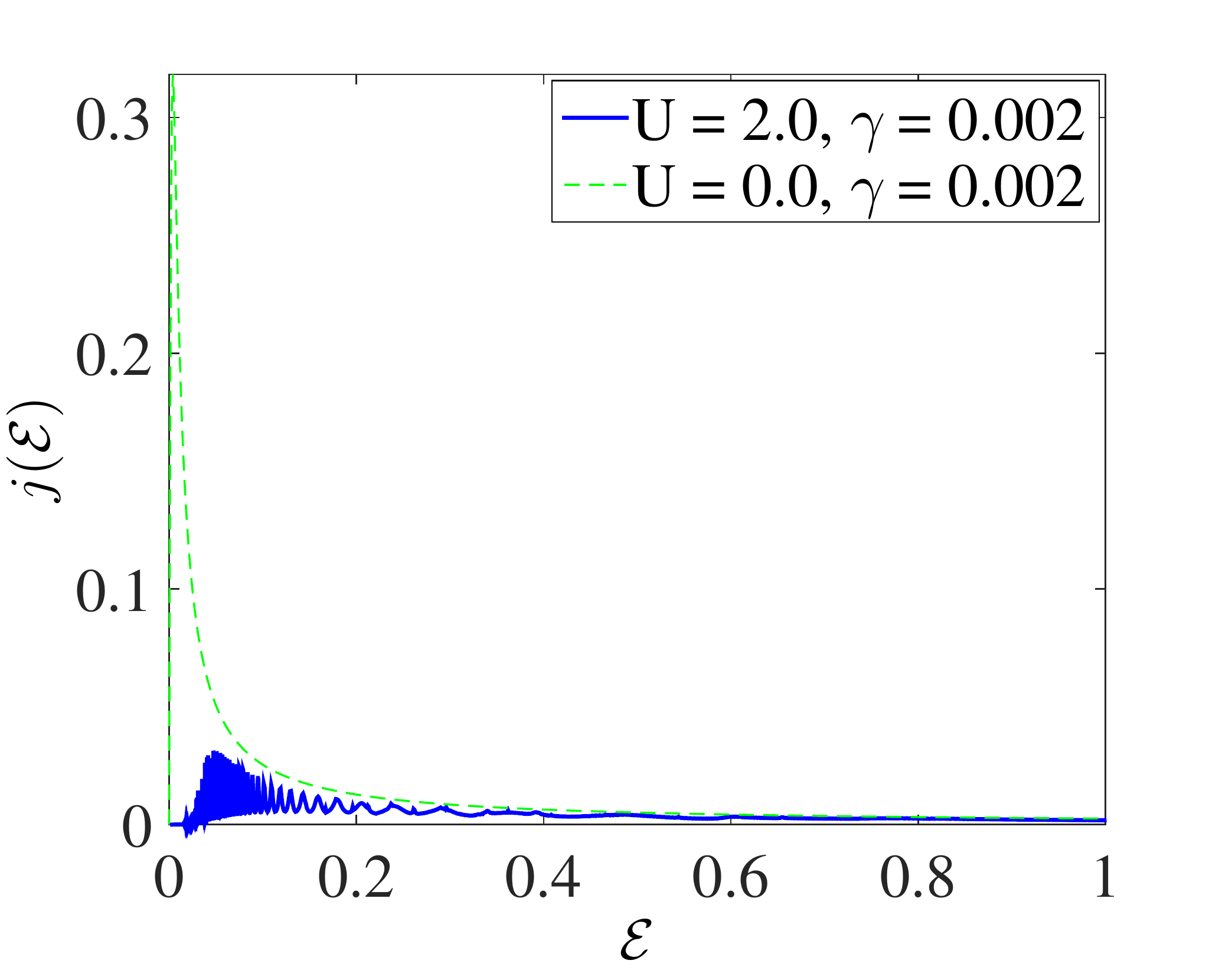}
\includegraphics[width=0.32\textwidth]{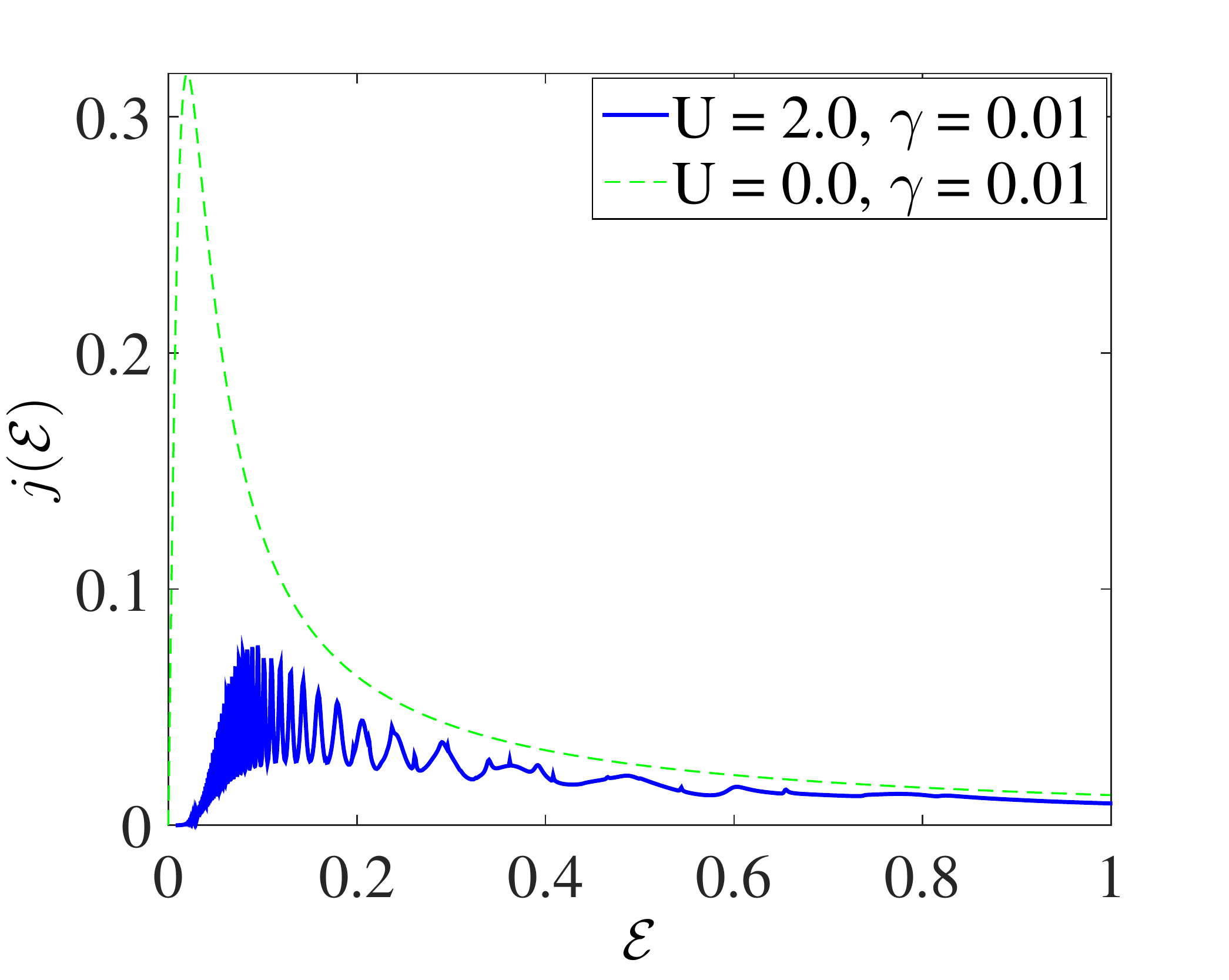}
\includegraphics[width=0.32\textwidth]{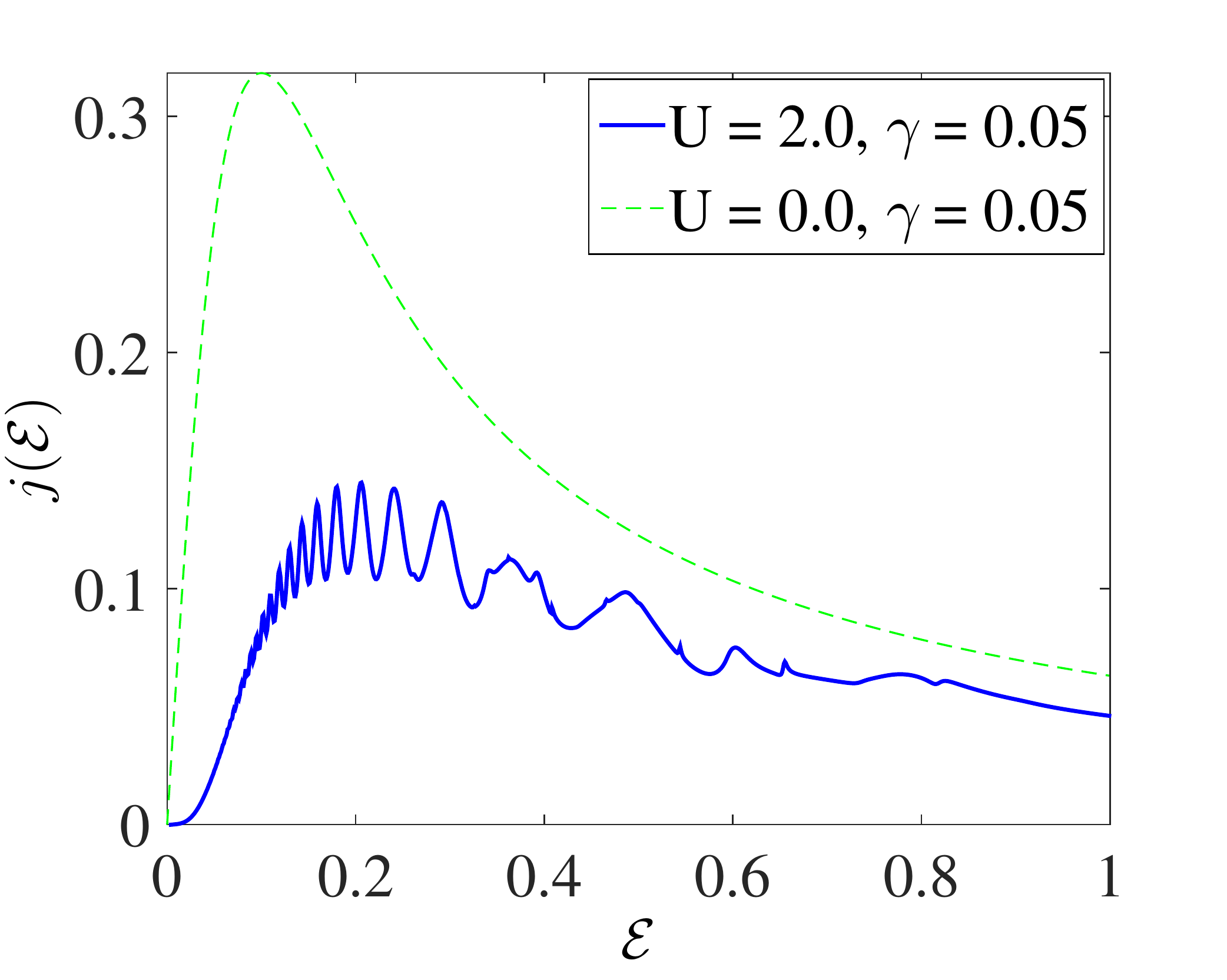}
\captionsetup{justification=raggedright,
singlelinecheck=false}
\caption{(Color online) Current $j(\EE)$ (blue solid line, calculated according to \eq{eq:CurrentDensity}) for  different values
of the effective damping $\gamma$ = 0.002, 0.01, 0.05  and 
a fixed value for the repulsive on-site energy $U = 2.0$. The corresponding $U=0$-result is shown as  reference 
(green dashed line).}
\label{fig:CurrentInteractingVarGamma}
\end{figure*}

\vspace{-0.5cm}

\eq{eq:j:han} emphasizes the importance of the heat-bath.
Without bath chains ($\gamma$=0) the resulting current would be zero for all values of $\EE$, as it was also observed in recent studies for the 1D\cite{han.13} and 2D\cite{am.we.2012} Hubbard model.

The linear relationship $j(\EE) \propto \EE$ valid for weak electric fields, corresponds to Ohm's Law formulated in the Drude model:
\begin{align}
\label{eq:gradient}
\left. \frac{j}{\EE} \right|_{\EE \ll \gamma} & \approx \frac{t}{ \pi\,\Gamma} \propto \frac{e^2 \tau}{m^*}\, \mbox{,}
\end{align}
where $\tau$ denotes the lifetime of a transport electron with an effective mass $m^*$ and a charge $e$ and the TB relationship $m^* \propto 
\frac{1}{t}$ and $\gamma \propto \frac{1}{\tau}$ was used.\cite{omar.1975, gl.ko.2006} The expression essentially represents the well-known 
Drude DC conductivity per electron.

The initial linear region agrees with that obtained by linear response theory.\cite{kubo.57} The occurrence of a maximum value 
for the current in combination with a subsequent negative differential conductance 
can be explained by the increasing relevance of BOs at larger electric field intensities:
When the electron moves to a neighbouring site of lower potential energy, its kinetic energy 
\textit{increases} by $\EE$ (in the present units). At the same time 
part of the energy is \textit{dissipated} into the heat-bath chain. The
energy loss due to \textit{dissipation} is given in terms of the effective damping $\gamma$. For $\EE \gg \gamma$ 
the dissipation mechanism is simply too weak to dissipate enough energy and consequently prevent the occurrence of BOs, 
which results in a continuously decreasing current. Similar results have been reported  recently  for the 
2D Hubbard model.\cite{am.we.2012}

\subsection{Current in the Hubbard-Wannier-Stark model}\label{ssec:Interacting}

Next we want to study the impact of electronic correlations. 
For $U\ne 0$ the cluster size $M$ becomes important. 
We start the following investigations with  a 
two-site cluster ($M=2$). We will also present the results
for larger cluster sizes ($M=4$ and $M=6$) before turning back to the two-site cluster treatment to
explain the origin of the arising current resonances in the Hubbard-Wannier-Stark model.

\Fig{fig:CurrentInteracting} shows the dependence of the current $j(\EE)$
on the interaction strength. In all cases we used a damping parameter $\gamma = 0.01$.
For $U\le 0.5$ we find similar  $j(\EE)$-curves as for $U=0$, where the current maximum 
decreases slightly with increasing $U$. For $U\ge 1$ an oscillatory behaviour sets
in and a transport gap opens up. This gap is however much smaller then the
single-particle gap that one obtains for $\EE=0$ in the density of states.

Similar resonances have been found experimentally  in 
$\text{GaAs/Al}_{x} \text{Ga}_{1-x} As$ superlattices.\cite{andronov_transport_2008}
The origin of the resonant structures in the current 
visible in \Fig{fig:CurrentInteracting}
will be discussed in \se~\ref{ssec:ResonantTunneling}. 

First, however, we will study the effect of the dissipation strength on the current characteristics, which are displayed in
\Fig{fig:CurrentInteractingVarGamma} for a fixed value of $U = 2.0$ 
and various values of $\gamma$.
There are two major effects observable in \Fig{fig:CurrentInteractingVarGamma}. First of all, 
the current maximum $j_\text{max}$ increases monotonically with increasing dissipation strength. 
This is in contrast to the result at $U=0$, where only the position of the maximum current is 
shifted according to \eq{eq:j:han:max}.
But it corroborates that a non-zero DC current is only possible when  dissipation  is included. 
 \begin{figure*}[t!]
 \includegraphics[width=0.32\textwidth]{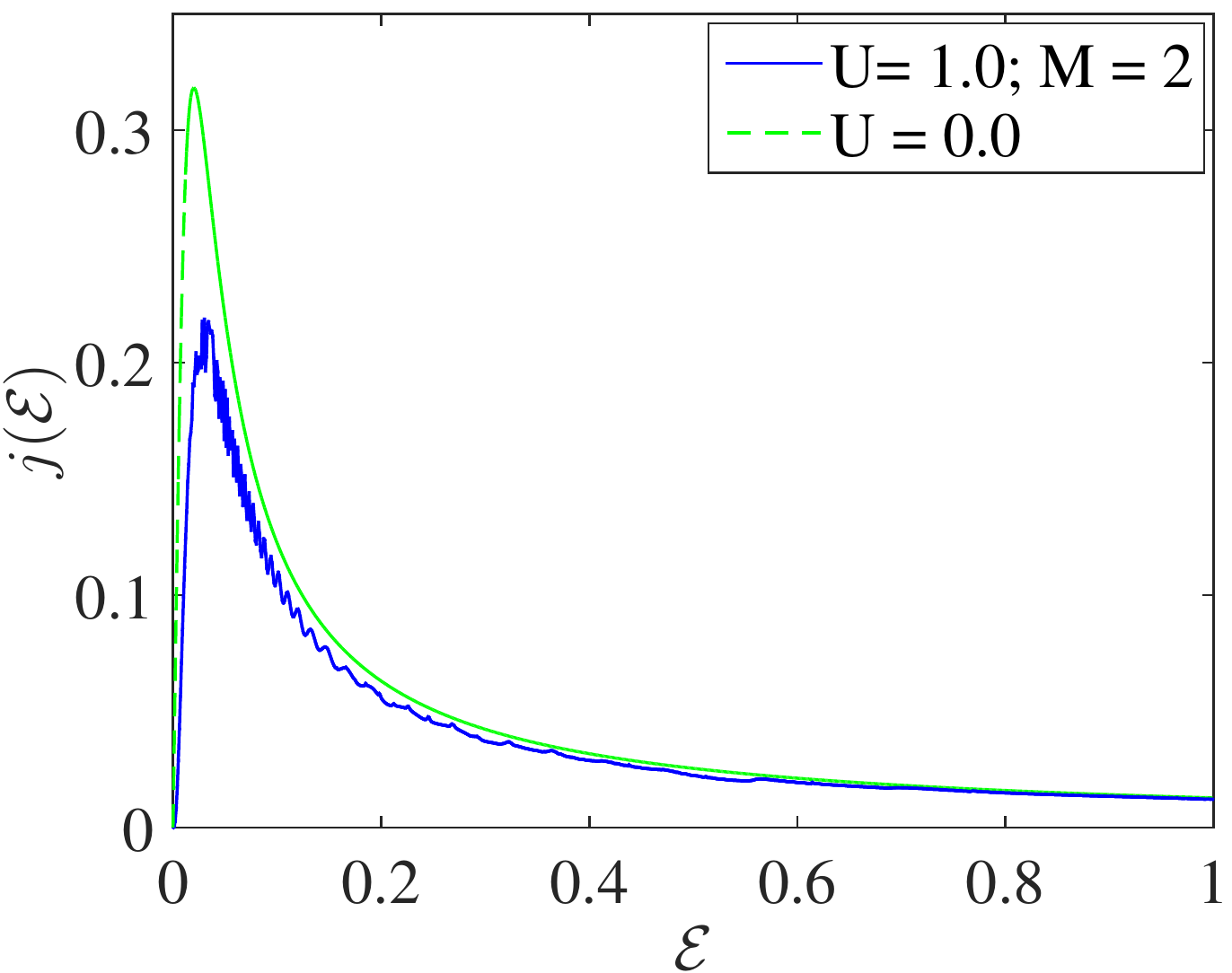}
\includegraphics[width=0.32\textwidth]{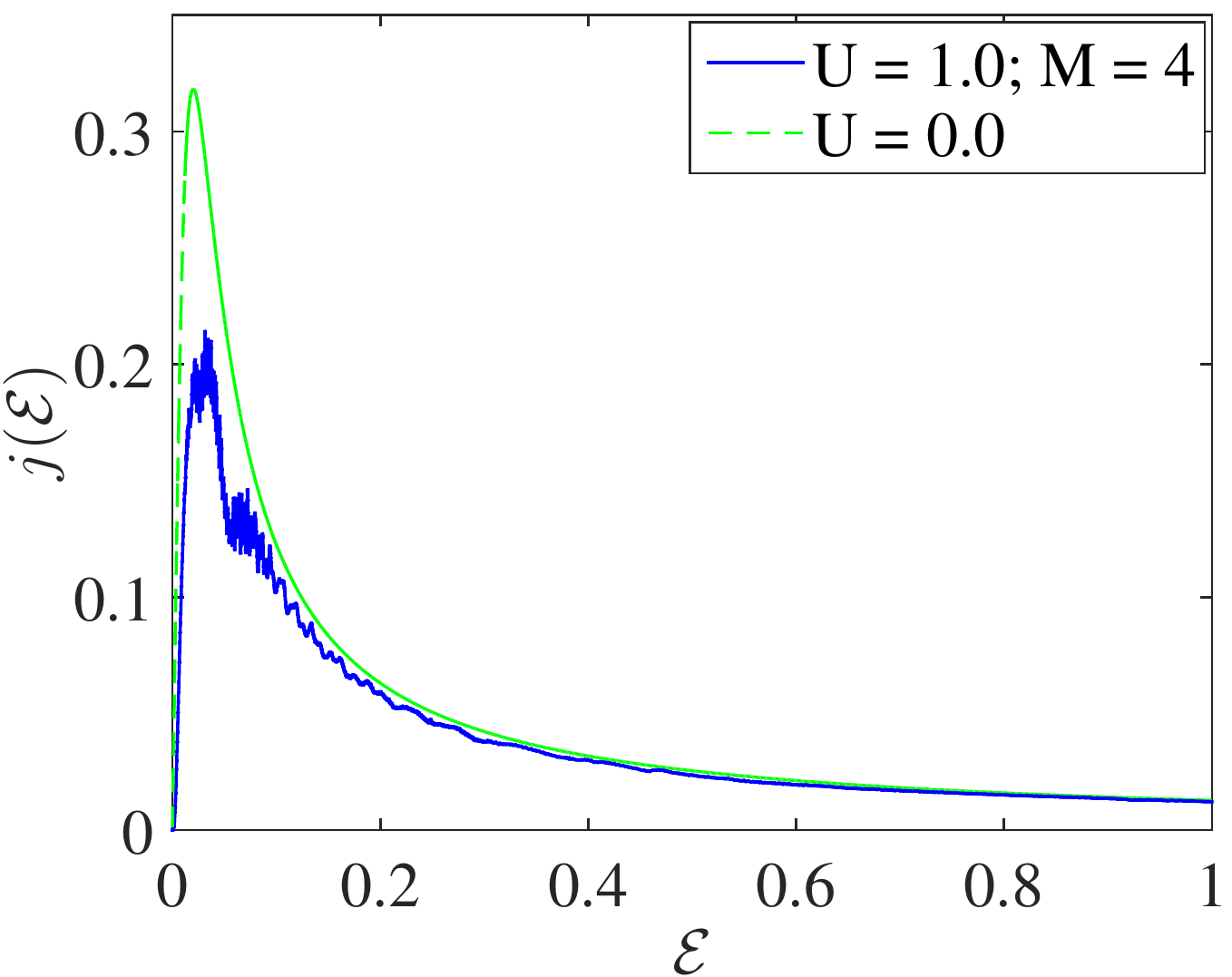}
\includegraphics[width=0.32\textwidth]{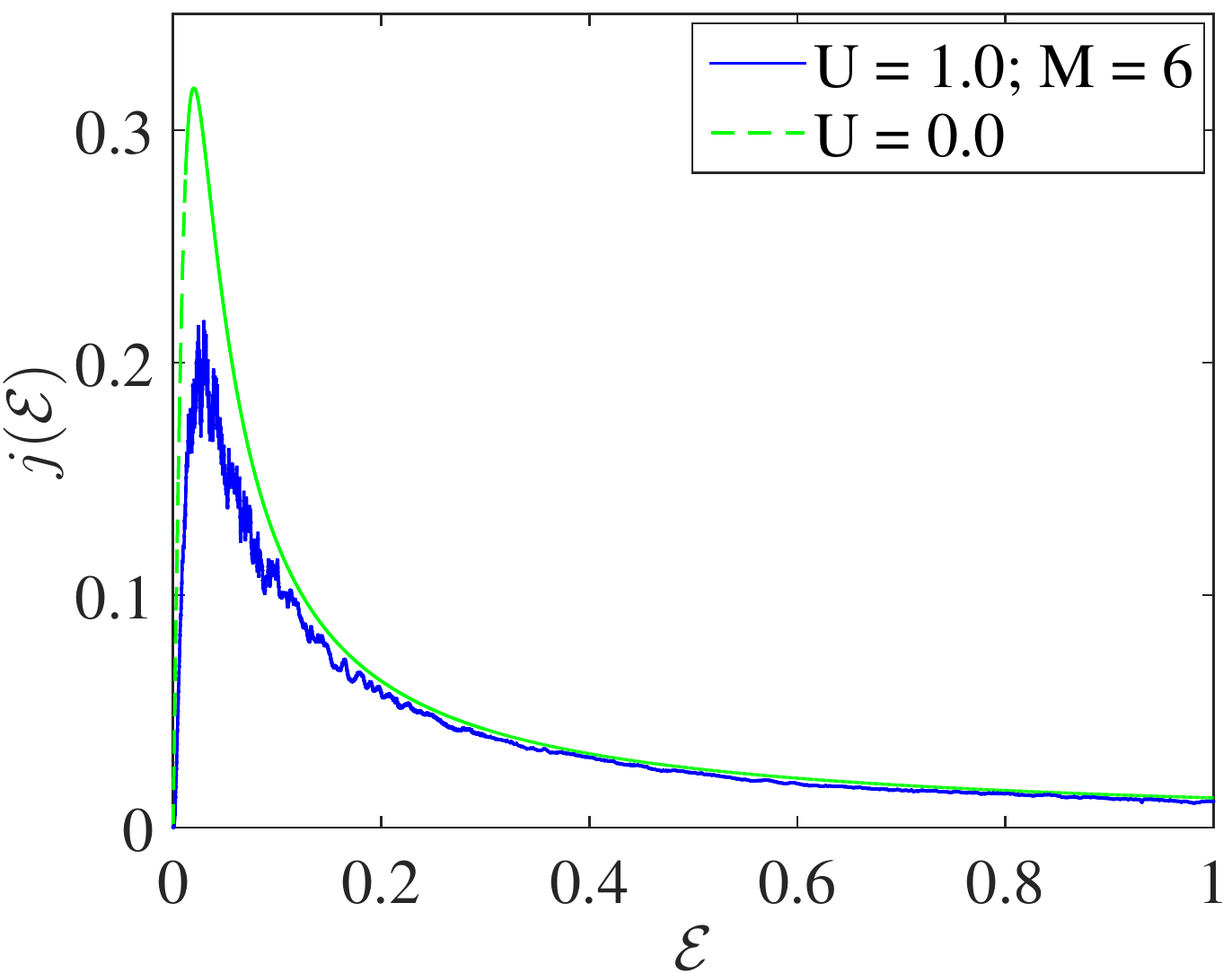} \\ \vspace{0.1cm}
 \includegraphics[width=0.32\textwidth]{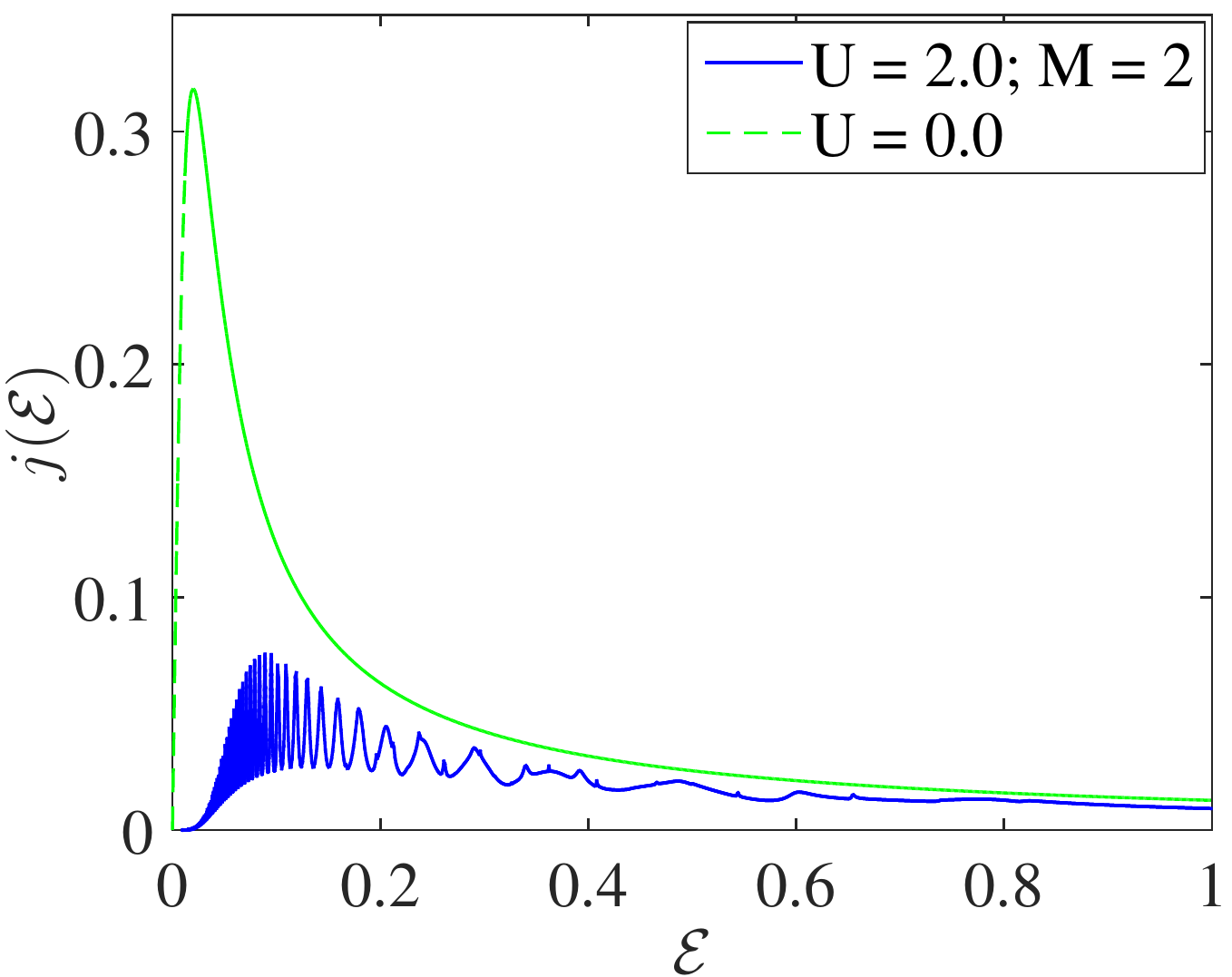}
\includegraphics[width=0.32\textwidth]{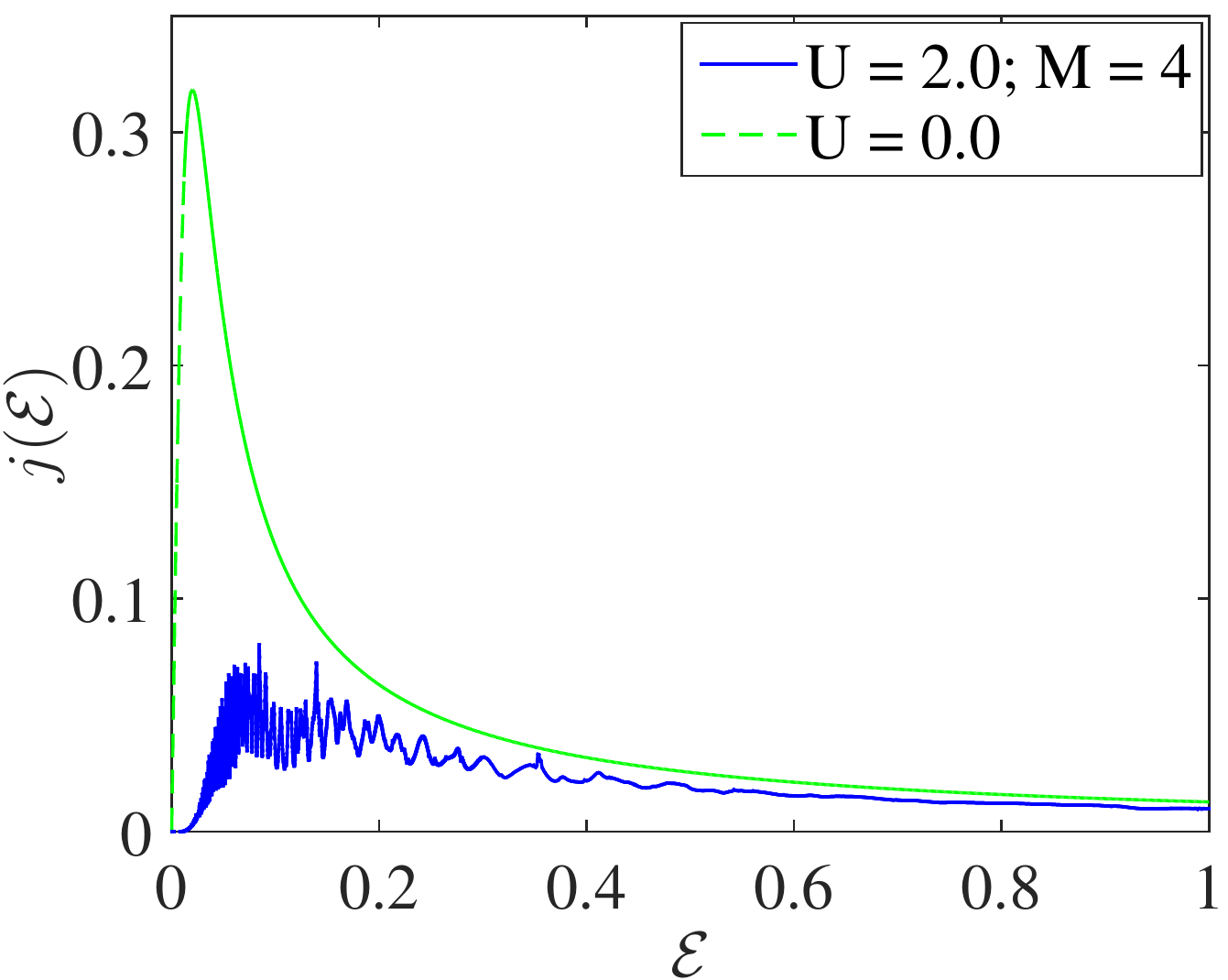}
\includegraphics[width=0.32\textwidth]{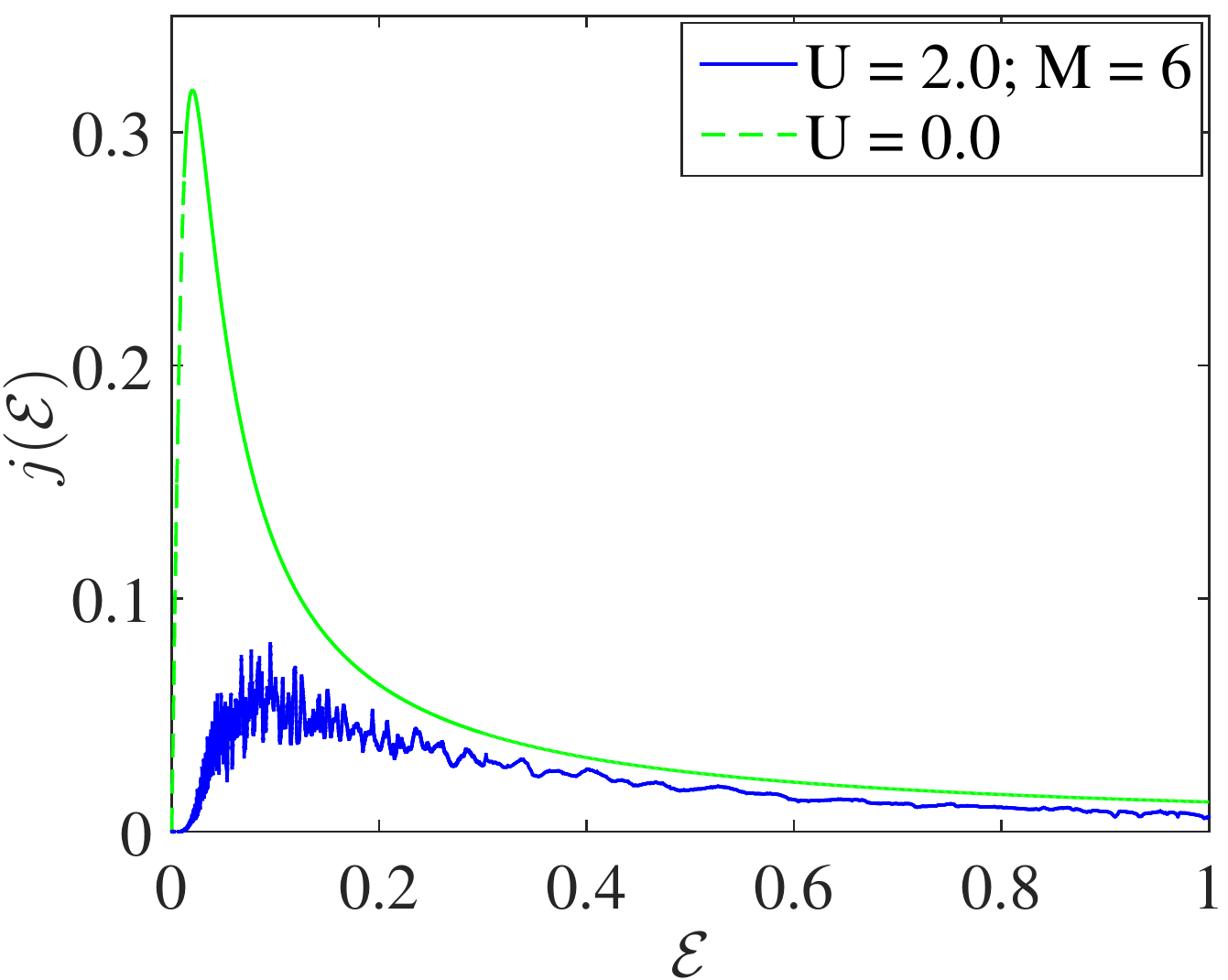}
\captionsetup{justification=raggedright,
singlelinecheck=false}
\caption{(Color online) Current $j(\EE)$ (blue solid line, calculated according to \eq{eq:CurrentDensity}) for  different cluster
sizes $M$ = 2, 4, 6. The results are shown for an interaction strength $U = 1.0$ (top row) and $U = 2.0$ (bottom row) with a fixed 
value of the effective damping $\gamma = 0.01$. The corresponding $U=0$ result is shown as reference (green dashed line).}
\label{fig:CurrentInteractingVarSizes}
\end{figure*}
%
The second effect is a smoothing of the resonances with increasing dissipation strength. While one finds a very spiky 
structure in the $j(\EE)$ curve for a small effective damping, the oscillations for larger values of $\gamma$ are rather smooth. 
The origin of this behaviour will be discussed in the following
section.
\vspace{-0.08cm}

Now we focus on the influence of
larger cluster sizes on the current characteristics analyzed in \Fig{fig:CurrentInteractingVarSizes}. 
Independent of the interaction strength the qualitative behavivour does not change significantly when the number 
of sites within the CPT cluster is increased. The spiky structure of the current remains as well as the finite transport gap.
Furthermore, comparing the results for $M=4$ and $M=6$ it seems that in a wide region of the electric field the results coincide 
also in a quantitative way. Due to the rather weak impact of the cluster size we will restrict our further investigations 
to the $M=2$ case. 

\vspace{-0.5cm}
\subsection{Origin of the resonances in the Hubbard-Wannier-Stark model}\label{ssec:ResonantTunneling}

We will argue in this section that the origin of the resonant structures can be traced back to the 
occurence of short-range antiferromagnetic order. Although it is well known that the 1D Hubbard model in equilibrium
does not exhibit long-range order,\cite{MerminWagner,da.ai.04} strong short-range fluctuations are nevertheless present.
On the short length scale of the cluster, where true long-range and short-range order can not be distinguished, 
this order can be described reasonably well by a mean-field decoupling,
i.e. by the Stoner model. We will show in the following, that the positions of the resonances are in perfect agreement 
with the maxima of the eigenvalues of the WSL obtained in this mean-field treatment. 

The mean-field decoupled Hamiltonian is given by
\begin{align}
 \label{eq:H_IA_MF}
\hat{\mathcal{H}}_{1} \approx \hat{\mathcal{H}}_{1}^{\text{MF}} &= \frac{U}{2} \, \sum\limits_{j\sigma} 
\big\langle\big({n}_{j\overline{\sigma}} -\frac{1}{2}\big)\big\rangle
\big({n}_{j\sigma} -\frac{1}{2}\big)
\;,
\end{align}

\begin{center}
 \begin{figure*}[t!]
\subfloat{
\includegraphics[width=\textwidth]{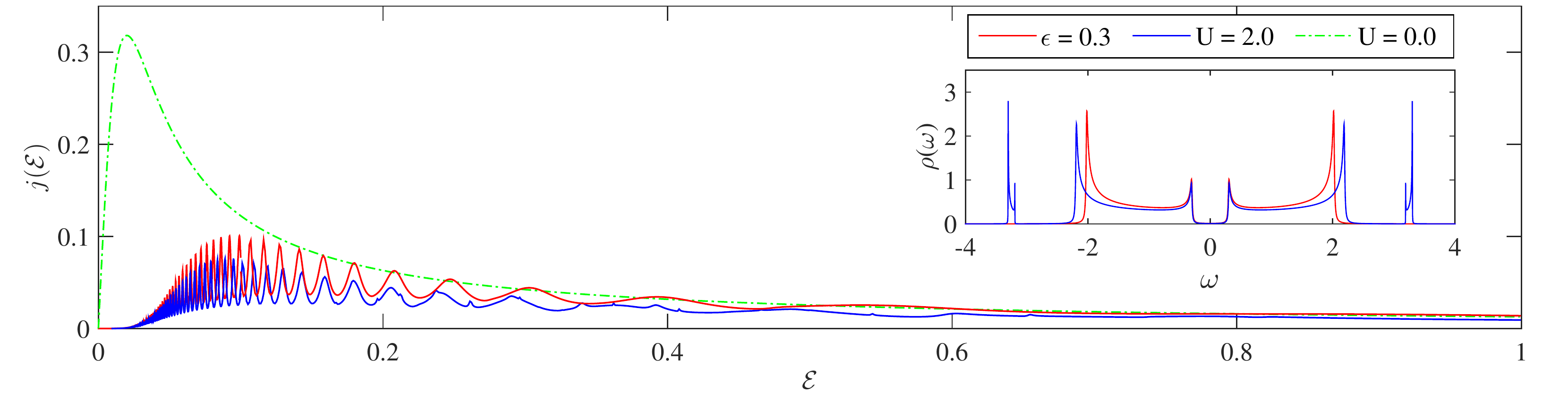}}  \\ \vspace{-0.024\textwidth}
\captionsetup{justification=raggedright,singlelinecheck=false}
\caption{(Color online) Comparison of the current characteristics $j(\EE)$ obtained for the
Hubbard model  and its mean-field approximation, both for $\gamma$ = 0.01.   
For the Hubbard model $U=2.0$ and in the mean-field approximation the corresponding
order parameter is $\epsilon = 0.3$. The latter is
adjusted such that both models have the same 
single-particle energy gap,
as can be seen in the inset, which depicts the corresponding equilibrium DOS.}
\label{fig:ModelComparison}
\end{figure*}
\end{center}

Assuming anti-ferromagnetic order leads to 
\begin{align}
 \label{eq:H_IA_MF:2}
 \hat{\mathcal{H}}_{1}^{\text{MF}} &= \epsilon \, \sum\limits_{j\sigma} 
e^{i \pi j}
\big({n}_{j\sigma} -\frac{1}{2}\big)
\;,
\end{align}

where the value for the order parameter $\epsilon$ will be adjusted such that the Stoner
model gives the same single-particle energy gap as the Hubbard model.
The inset of \Fig{fig:ModelComparison} shows the  equilibrium Density of States (DOS) 
$\rho(\omega)$ for the Hubbard and the Stoner model. Obviously, the gap and the low-lying excitations are well 
described by the Stoner model.

More importantly, we see that the current characteristics 
of the two models are qualitatively in very good agreement. Concerning the position of the resonances, 
we see even perfect quantitative agreement. This allows a transparent 
explanation of the resonances in the frame of the much simpler Stoner model. To this end we start out with the bare Stoner model ($\hat{\mathcal{H}}^\text{WSL} =  \hat{\mathcal{H}}_{0} + \hat{\mathcal{H}}_{1}^\text{MF}$) without coupling to the bath chains.

\begin{figure}[ht!]
\includegraphics[width=0.35\textwidth]{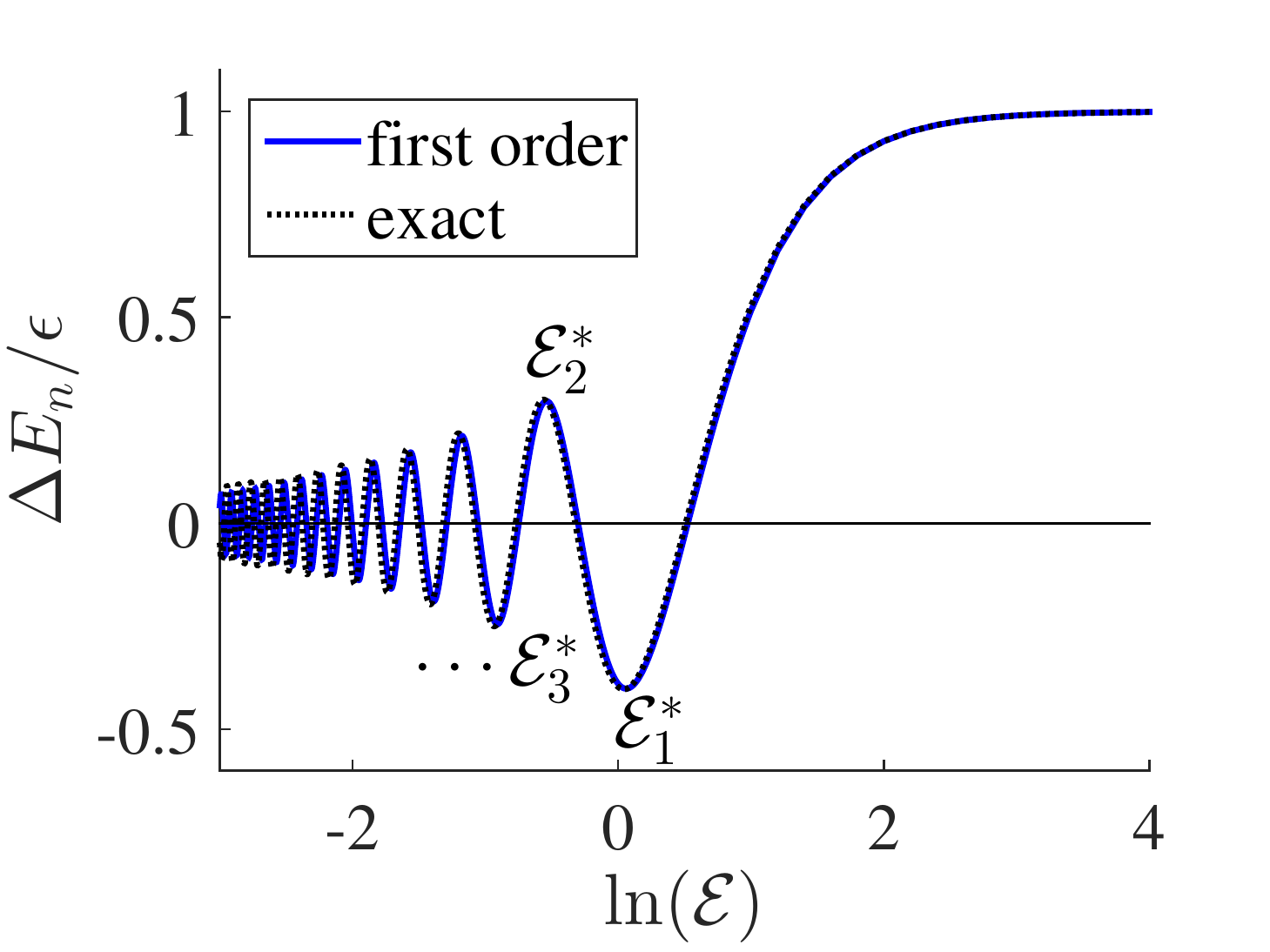}
\caption{(Color online) 
Energy correction $\Delta E_{n} = E_{n} - n \EE$ for  even $n$ as function of the logarithmic 
field strength. A comparison is given between the first order correction and the exact value, which is obtained numerically by exact diagonalization. Note that $\Delta E_{n}$ it is the same for all even sites and for odd $n$ the sign is reversed.
}
\label{fig:Delta:E}
\end{figure}

\begin{figure}[ht!]
\includegraphics[width=0.5\textwidth]{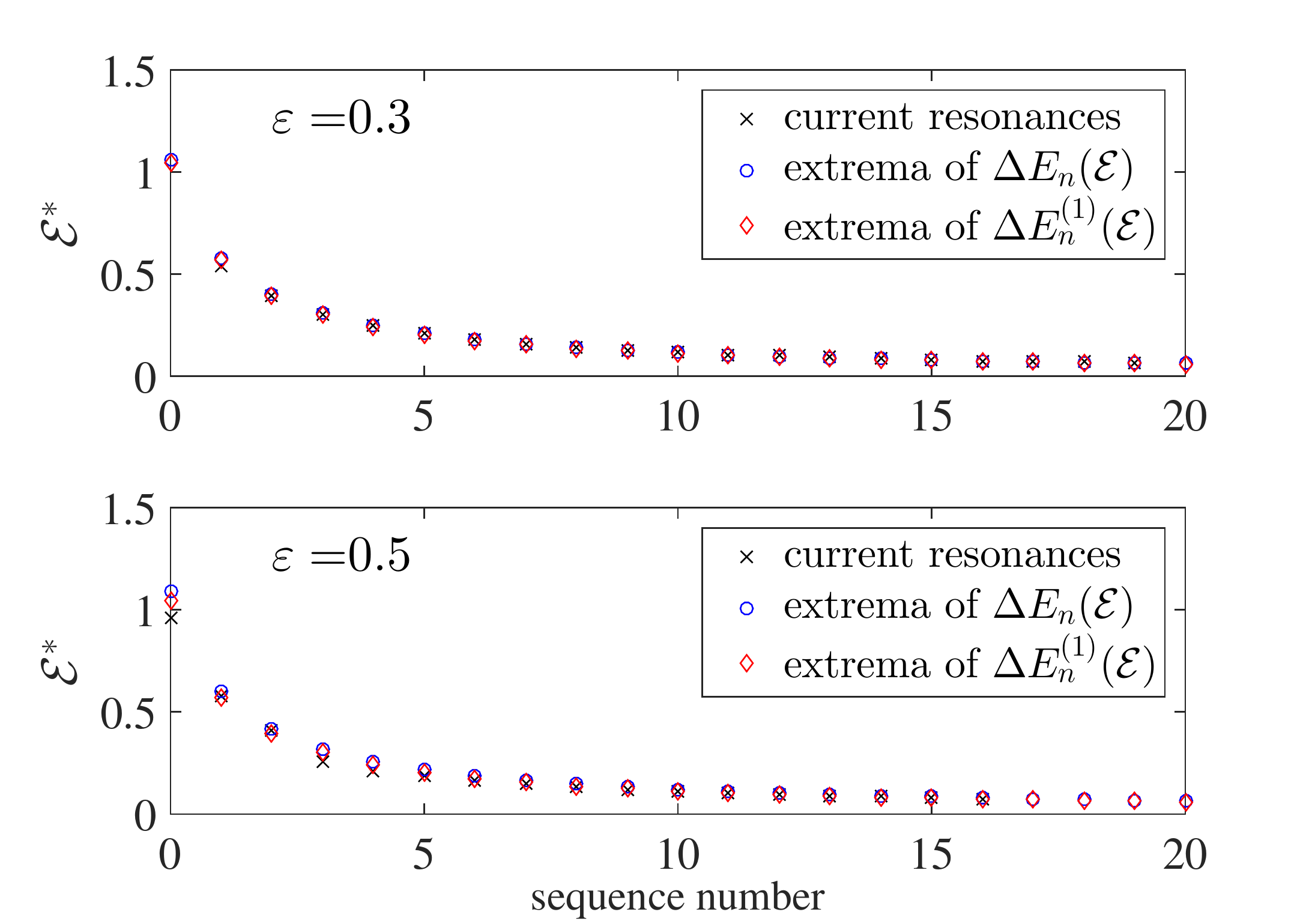}
\caption{(Color online) Field values corresponding to the current resonances for the Stoner model, depicted in \fig{fig:ModelComparison}, compared with the positions of 
extrema of $\Delta E_{n}(\EE)$ for the exact eigenvalues 
and those obtained by  first order perturbation theory, as discussed in the text.}
\label{fig:resonances}
\end{figure}

It is well known (see e.g. ref. \onlinecite{ha.mo.04}) that non-interacting electrons in a 
periodic potential, which experience in addition a homogeneous electric field, are trapped in localized states, corresponding
to Bloch-oscillators. The energies form \textit{Wannier-Stark ladders} (WSL). In the present example of an alternating potential
the WSL-energies  have the form

\begin{align}\label{eq:WSL:Delta}
E_{n} &= n \EE + \Delta E_n\;,
\end{align}
where $\Delta E_n$ denotes a possible energy correction.
Without the periodic potential ($\epsilon=0$) the WSL-energies  are simply 
$E_{n} = n\EE$, corresponding to $\Delta E_n=0$. 
It is straightforward to determine the eigenvalues (and hence $\Delta E_n$)
of $\hat{\mathcal{H}}^\text{WSL}$ numerically for sufficiently large systems.
The result  is depicted in \fig{fig:Delta:E} for $\epsilon=0.5$ as function of $\ln(\EE)$.
We observe that  for $\EE\gg 1$ the energy correction $\Delta E_n$ approaches the value
$\epsilon$, the amplitude of the alternating potential. For $\EE< 1$ the energy correction
oscillates with extrema at $\EE^{*}_{\nu}$ as indicated in \fig{fig:Delta:E}.
Very surprisingly, it turns out
that these field strengths correspond to the values of $\EE$ where the current maxima are observed in
\fig{fig:ModelComparison}. This is illustrated in  \fig{fig:resonances}
where the extrema of $\Delta E_{n}$ are compared with the position of the current resonances.

In order to unravel the origin of these oscillations in $\Delta E_n(\EE)$ it is 
instructive to solve the eigenvalue problem in first order perturbation theory. 

The eigenvectors  of $\hat{\mathcal{H}_{0}}$ can be expressed as\cite{ha.mo.04} 

\begin{align}\label{eq:WSL:states}
\ket{\Psi_m} &= \sum_{n} 
J_{n-m}(\zeta)\;
 c^{\dagger}_{n} \ket{0}\;,
\end{align}

where $J_{l}(\zeta)$ are Bessel functions of the first kind and
 $\zeta=\frac{2t}{\EE}$. 
 In \app{app:perturbationTheory} we show that the first order energy corrections 
are given by

\begin{align*}
\Delta E^{(1)}_{m} = \bra{\Psi_{m}} \hat{\mathcal{H}^\text{MF}_{1}}\ket{\Psi_{m}}
&= e^{i\pi m} \epsilon \;J_{0}(2\zeta)\;.
\end{align*}


The first order correction is also plotted in \fig{fig:Delta:E} and compared with the exact correction
$\Delta E_{n} := E_{n}- n \EE$.
Obviously, the energy correction is well described by the first order term and it also 
shows the oscillatory dependence on $\EE$. To understand this behaviour it is crucial to bear in 
mind that the \textit{Wannier-Stark states} $\ket{\Psi_{n}}$ are localized with mean position   
$\avg{m}=n$ and variance 

\begin{align}\label{eq:WSL:sigma}
\sigma^{2}:=\avg{(m-\avg{m})^{2}} = \frac{\zeta^{2}}{2}=\frac{2t^{2}}{\EE^{2}}\;.
\end{align}
For completeness the proof is given in \app{app:WannierStarkLadder}.

Consequently, for fields $\EE> \sqrt{2} $ the state is confined to a single site, $n=0$ 
say, and the first order energy correction is simply $\Delta E_0^{(1)} = \epsilon$, corroborating 
the asymptotic behaviour in \fig{fig:Delta:E} for $\EE\gg 1$.
Then,  upon lowering $\EE$ the wavefunction spreads out like $1/\EE$  and reaches gradually 
sites with alternating potential, which leads to an oscillatory behaviour of $\Delta E^{(1)}$ as 
observed in \fig{fig:Delta:E}. 

In perturbation theory, we can also determine the position of the extrema of $\Delta E_n^{(1)}(\EE)$
analytically 

\begin{align*}
\frac{\partial }{\partial \EE} \Delta E^{(1)}_{m}(\EE)\big|_{\EE=\EE^{*}}  
&\propto 
\frac{\partial }{\partial \EE} J_{0}(\frac{4t}{\EE})\big|_{\EE=\EE^{*}}
\propto J_{1}(\frac{4t}{\EE^{*}})
\overset{!}{=}0\;.
\end{align*}

Let $\xi^{(1)}_{\nu}$ be the $\nu$-th zero of $J_{1}(x)$ then

\begin{align}\label{eq:EEs:pt}
\EE^{*} &= \frac{4t}{\xi^{(1)}_{\nu}}\;.
\end{align}

Interestingly, to first order, the positions of the maxima are independent of $\epsilon$. 
In \fig{fig:resonances} these positions are also included. 
We find that the first order result is in reasonable agreement with the current resonances.
\begin{center}
\begin{figure}[b!]
\includegraphics[width=0.35\textwidth]{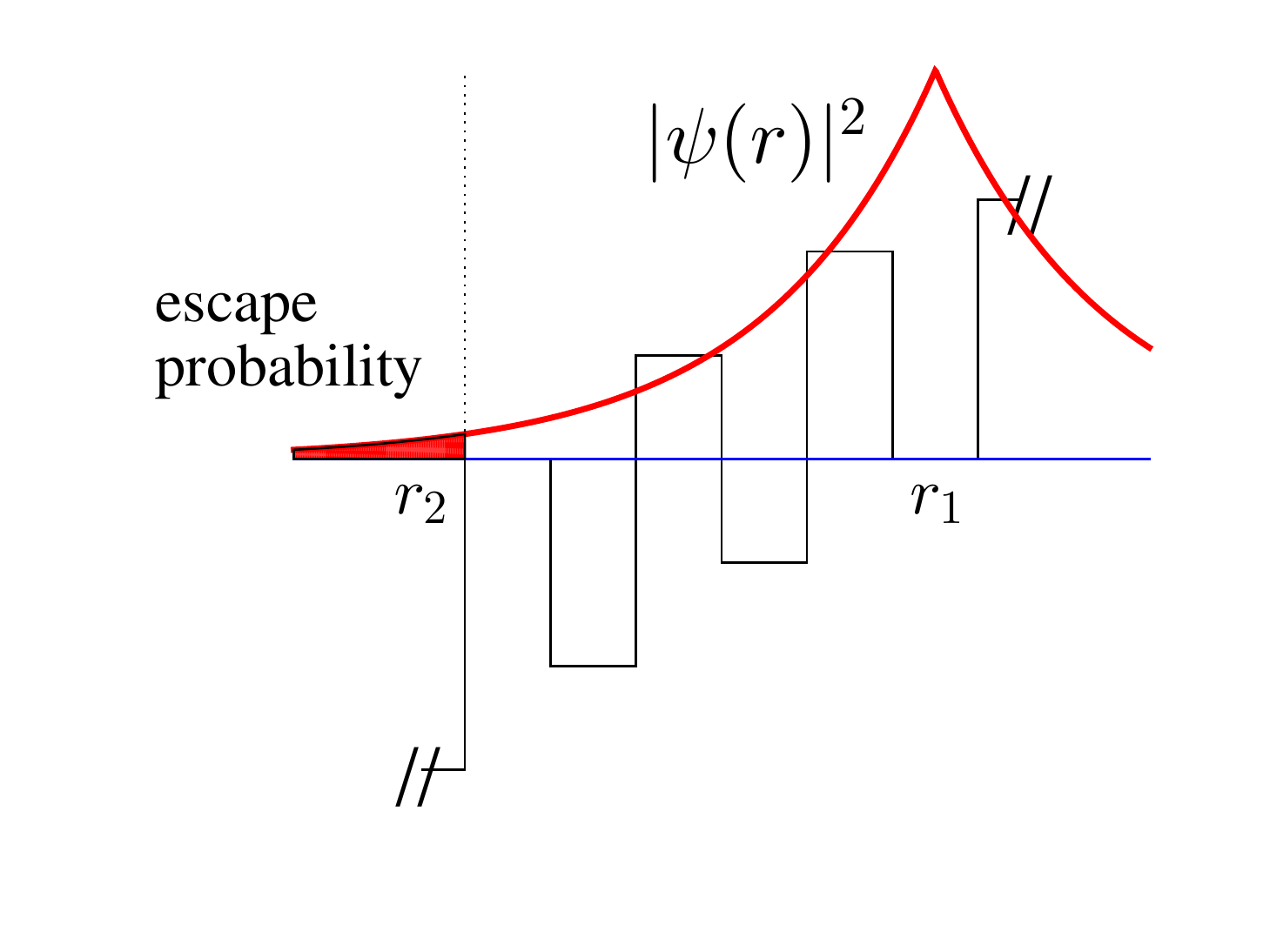}
\caption{(Color online) On-site energies (solid black line) of the Stoner model in a homogeneous electric field with anti-ferromagnetic order parameter.
The red curve sketches the wavefunction localized in the potential well at site $r_{1}$. The potential well ends on the left site  at position $r_{2}$. The escape probability is proportional to the size of  the filled red  area.}
\label{fig:PotentialTunnel}
\end{figure}
\end{center}
\vspace{-0.7cm}
We can now invoke second order Fermi's golden rule to compute the current induced by the
coupling to the bath chains. The first order terms vanish as the coupling to the bath changes
the number of electrons in the physical chain. Brute force application of Fermi's golden rule
shows that the energy correction $\Delta E_n(\EE)$ play a decisive role and 
Fermi's golden rule corroborates the above observation that the current maxima occur at the extrema of $\Delta E_n(\EE)$.

Now that we have shown that the short-range order is responsible for the current resonances, 
we want to analyze the
current suppression for small $\EE$ observed in \fig{fig:ModelComparison}.

For small $\EE$ the Wannier-Stark states are the wrong starting point. In this case we have to consider $\EE$ as perturbation. Then the localized states are due to the alternating potential barrier and have an $\EE$-independent localization width. 
In \fig{fig:PotentialTunnel} a sketch of the localized wavefunction 
centred at some site $r_{1}$ is depicted. The $\EE$-field induced linear potential allows the particle to tunnel through
the potential barrier that ends at site  $r_{2}$. The escape probability is proportional
to the size of the filled area in  \fig{fig:PotentialTunnel}. 

For a qualitative description we approximate the left half of the wave function 
by an exponentially decreasing function
\begin{align}
\label{eq:Wavefunction}
\psi(r) & \propto \exp\left(-\frac{|r-r_{1}|}{\xi}\right) \;.
\end{align}
The quantity $\xi$ describes the spatial extent of the localized wavefunction.
\begin{center}
\begin{figure}[h!]
\subfloat{\label{}
\includegraphics[width=0.3\textwidth]{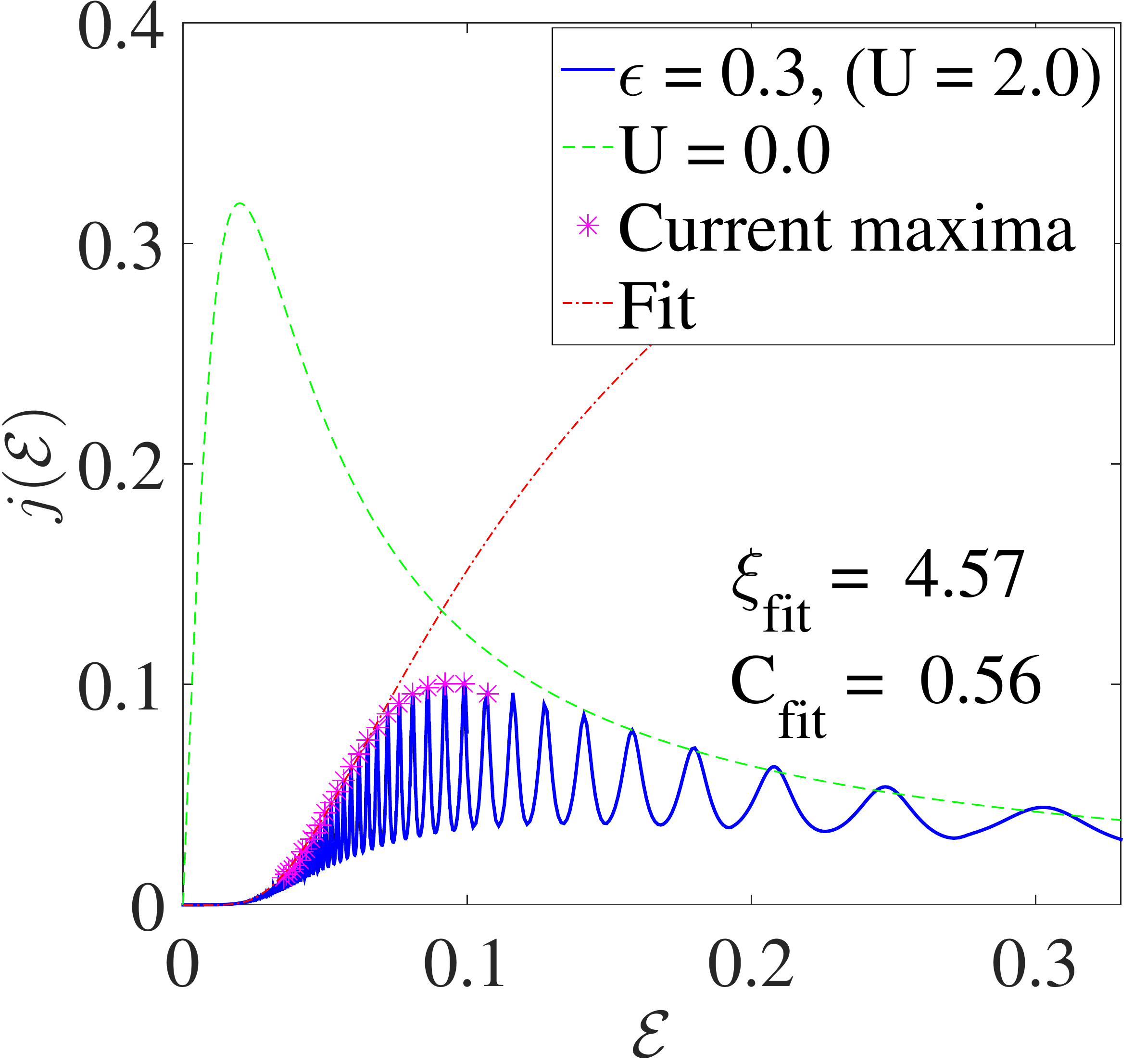}} 
\captionsetup{justification=raggedright,singlelinecheck=false}
\caption{(Color online) Plot of \Fig{fig:ModelComparison} zoomed into the low electric field region.
The maxima are highlighted (magenta star markers) and fitted according to \eq{eq:Res1} 
(red dashed-dotted lines) resulting in the  fit parameters $\xi_{\text{fit}}$, $C_{\text{fit}}$,
depicted in the figures.
}
\label{fig:Fits1}
\end{figure}
\end{center}
\vspace{-0.1cm}
Then the escape probability and in turn the current should be proportional to
\begin{align}
\label{eq:FiniteProbability}
j(\EE) \propto \int_{-\infty}^{r_2} \left|\psi(r)\right|^2 dr 
&\propto \exp\left(-\frac{2(r_{1}-r)}{\xi}\right)  \;.
\end{align}
The distance $r_{2}-r_{1}$ can be approximated by the slope of the linear potential 
$r_{2}-r_{1}\approx 2\epsilon/ \EE$. Hence, we have

\begin{align}\label{eq:Res1}
j(\EE) &= C \exp\left(-\frac{4 \epsilon}{\EE\cdot\xi}\right) \;.
\end{align}
We consider $\xi$ as fit parameter, which is estimated from the height of the resonances
for small  electric fields.
The resulting envelope  is depicted in \fig{fig:Fits1}.
The envelope function describes the values of the current maxima almost perfectly 
for small $\EE$, which confirms our description. 
The size of the transport gap, defined by the field strength $\EE^{*}$ for which
the exponential in \eq{eq:Res1} is unity, is characterized by
\begin{align}\label{eq:transport:gap}
\EE^{*} &= \frac{4 \epsilon}{\xi}\;.
\end{align}
It is proportional to the height potential barrier and inverse proportional to the localization
length.
For large $\EE$ the current is suppressed 
by the BOs in agreement with the non-interacting result, which is also shown in \fig{fig:Fits1}.

\vspace{-0.2cm}
\subsection{Consequences/Relevance of non-local effects in the self-energy}\label{ssec:VCA}

\begin{figure}[h!]
\includegraphics[width=0.23\textwidth]{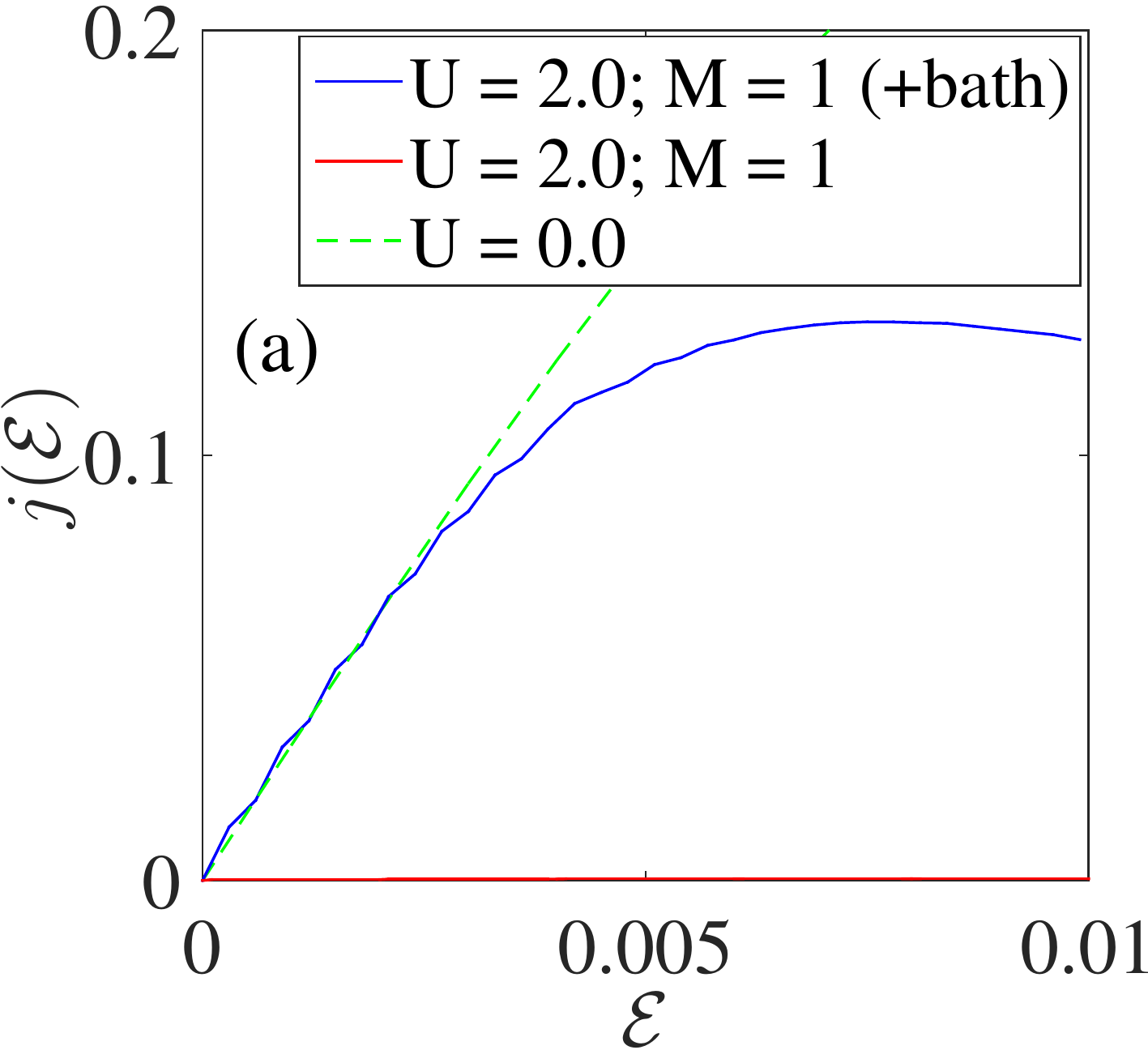}
\includegraphics[width=0.23\textwidth]{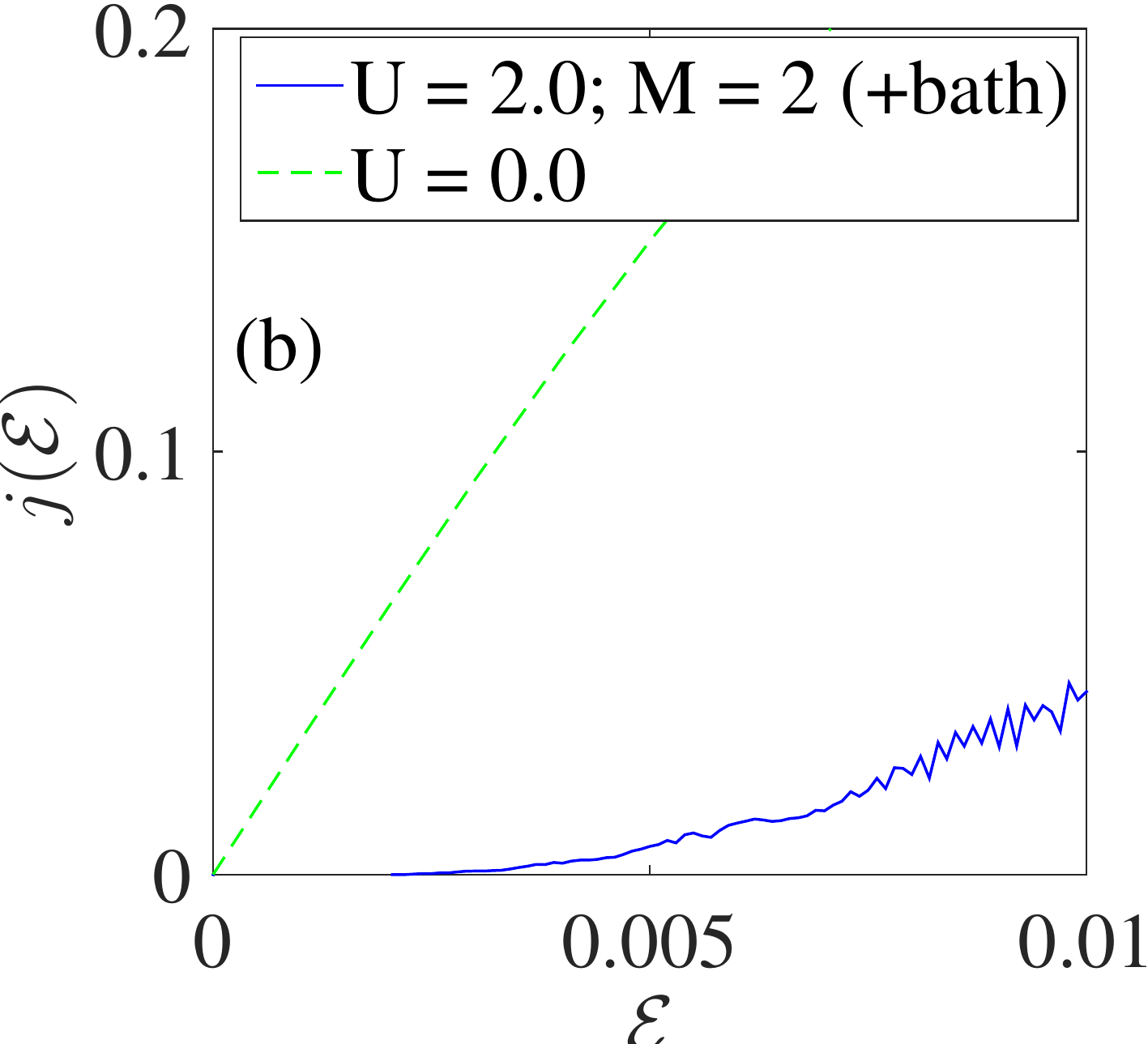}
\captionsetup{justification=raggedright,singlelinecheck=false}
\caption{(Color online) Linear response regime/region of the current $j(\EE)$ (calculated according to 
\eq{eq:CurrentDensity}) with (blue solid line) and without (red solid line) an additional VCA optimized bath site for the
single-site case ($M = 1$) (a) and the 2-site case ($M = 2$) (b) for $U = 2$. In the single-site treatment the overall linear 
response behaviour changes from insulating to metallic by adding an additional bath site, while for the $M=2$ case 
the insulating behaviour remains. All results are shown for a fixed 
value of the effective damping $\gamma = 0.01$. The corresponding result at $U=0$ is shown as  reference (green dashed line).}
\label{fig:CurrentVCA}
\end{figure}

Our CPT results deviate
from a recent investigation of the same  system within DMFT.\cite{kotliar.han.2014}
This can be especially seen in 
the linear response behaviour: while our calculations suggest the 
infinite 1d chain to be in an insulating state, the DMFT calculations display a metallic linear response. 
This  insulating state with a too large gap is a known drawback of CPT 
also known from the Hubbard I  approximation
which  corresponds to using a single-site cluster
($M = 1$), as can be seen in \fig{fig:CurrentVCA}a (red curve). 
This drawback can be corrected by adding an additional uncorrelated bath site 
 in the sense
of a minimal DMFT setup. 
The parameters of the bath site can be determined, e.g.,  
within the  Variational Cluster 
Approach (VCA).\cite{VCA1, VCA2, VCA3}
\fig{fig:CurrentVCA}a (blue curve) shows the resulting current in the linear response region. Here, it is appropriate to
 use the parameters 
  determined for the equilibrium case $\EE=0$.
As one can see, the insulating 
behaviour has changed to a metallic one.
On the other hand, at half filling, the Hubbard model is expected to be insulating for any  $U>0$ both in one as well as in two 
dimensions.\cite{CDMFT,sc.ge.15}
This behavior is clearly missed when neglecting nonlocal effects in the self-energy as in 
single-site DMFT.
A minimal setup to account for the true physical
insulating solution down to $U=0$
consists of a two-site cluster ($M = 2$) with additional bath sites,
as can be seen from the linear response region displayed 
 in \fig{fig:CurrentVCA}b.
 These results point out the importance of non-local effects in the self energy, which are induced by short-range 
 antiferromagnetic fluctuations.
This is also confirmed
in an early VCA work\cite{VCA1} according to which it is more effective to include 
\textit{physical} sites in the cluster, thus making the self-energy more nonlocal, rather than bath sites, which emphasize its  
dynamical character.
This is the reason why
 we have chosen in this paper to use 
large  clusters rather than smaller ones with bath sites.
Nevertheless, it would be interesting in this context to investigate the same infinite 1d 
system also by means of a 
\textit{non-equilibrium} cellular DMFT (CDMFT)\cite{CDMFT} approach, in particular in a doped situation.

\section{Conclusions}\label{sec:conclusion}

We have seen that Hubbard chains in an electric field couped to heat-bath chains exhibit an oscillatory/ resonant behaviour in the current characteristics $j(\EE)$. The coupling to the heat-bath is essential for a non-vanishing steady-state current.
The origin of the oscillations has been traced back to anti-ferromagnetic fluctuations which 
we mimicked by the Stoner model.
Quite generally we have shown that non-interacting electrons in an alternating potential
permeated by a homogeneous electric field, i.e. for a Wannier-Stark model with alternating
on-site energies, the current exhibits an oscillatory behaviour as function of the field strength 
$\EE$, which is directly linked to similar oscillations in the spacing of the Wannier-Stark ladder.
For small field strength the current is suppressed due to localization in the periodic
potential barriers. For field strength $\EE$ much larger then the damping parameter $\Gamma$, the the current is again suppressed due to localization or rather Bloch-oscillations, which are 
a ubiquitous feature of the Wannier-Stark model without dissipation.

\begin{acknowledgments}
We acknowledge
discussions with Takashi Oka and Fabian Heidrich-Meisner. This work is partially supported by the Austrian Science Fund (FWF)
P24081 and P26508.
M.A. is supported by the START program of the Austrian Science Fund FWF, project Y746. 
\end{acknowledgments}

\appendix

\section{Equivalence of the Coulomb and  the temporal gauge in the Hubbard-Wannier-Stark model}\label{app:GaugeTransformation}

Here we will briefly outline a proof, which is  closely related to that given in Ref.~\onlinecite{graf_electromagnetic_1995}, 
that in the Hubbard-Wannier-Stark model an electric field
can either be described by a potential term, corresponding to the Coulomb gauge used in \eq{eq:H}, or in a 
time-dependent Peierl's phase, corresponding to the temporal gauge.\cite{han.13} 
The two approaches are equivalent in the sense that they differ by a   unitary transformation 
which leaves the Liouville equation invariant. As a consequence, the steady state current 
computed via

\begin{align}\label{app:current}
j\propto \Im\{t_{l+1,l}\cdot\text{tr}\big(\hat\rho \,c^{\dag}_{l+1,\sigma} c^{\nag}_{l,\sigma}\big)\}
\end{align}

will be the same.

To prove this statement, we start out with the Coulomb gauge, where the electric field adds a 
potential term 

\begin{align*}
\hat{\mathcal{H}}_{\EE} &= \EE \sum_{j} j\; (n_{j} + n^{b}_{j})
\end{align*}

to the Hamiltonian $\hat{\mathcal{H}}_{0}$ without $E$-field. The total Hamiltonian is $\hat{\mathcal{H}}=\hat{\mathcal{H}}_{0}+\hat{\mathcal{H}}_{\EE}  $, where the first term also contains the Hubbard interaction.
In the temporal gauge, on the other hand, there is no  $\EE$-dependent potential term, instead the 
nearest neighbour hopping parameter $t_{l+1,l}$  is modified  to

\begin{align*}
t_{l+1,l}\to \tilde t_{l+1,l} = t_{l+1,l}\,e^{i a q \EE t}\;.
\end{align*}

Let the  Hamiltonian in temporal gauge be denoted by $\tilde {\mathcal{H}}$. 
We define a time-dependent unitary operator 

\begin{align*}
U(\tau) &=e^{-i \hat {\mathcal{H}}_{\EE}\tau}\;.
\end{align*}

It can readily be seen that $\tilde {\mathcal{H}} = U^{\dagger} \hat{\mathcal{H}}_{0} U$, 
where the Hubbard part is not modified, as the unitary operator only contains density 
operators.
Let $\hat\rho(t)$ be the density operator defined for the 
system in Coulomb gauge with  the corresponding Liouville equation

\begin{align*}
\frac{d}{dt} \hat \rho &= -i\left[\hat{\mathcal{H}},\hat \rho\right] \;.
\end{align*}

Then the transformed density operator $\tilde \rho = U^\dagger \rho U$ 
fulfils 
\begin{align*}
\frac{d}{dt} \tilde \rho &= -i\left[\tilde {\mathcal{H}},\tilde \rho\right] \;,
\end{align*}
which is the Liouville equation in temporal gauge. 
The initial value for $t=0$ is the same for both representations, i.e. 
$\tilde \rho_{0} = \hat \rho_{0}$. 
It should be noted that the gauge invariance is violated if the bath chains do not
experience the same E-field potential as the corresponding physical site. The reason 
is  the coupling term between the physical sites and the thermal bath.
In view of the
unitary transformation it is obvious that 
the current defined in \eq{app:current} is the same in both representations,
$j_{l+1,l} =\tilde j_{l+1,l}$, since

\begin{align*}
 \Im\{t_{l+1,l}\cdot\text{tr}\big(\hat\rho \,c^{\dag}_{l+1,\sigma} c^{\nag}_{l,\sigma}\big)\}
&=
\Im\{\tilde t_{l+1,l}\cdot\text{tr}\big(\tilde \rho \,c^{\dag}_{l+1,\sigma} c^{\nag}_{l,\sigma}\big)\}\;.
\end{align*}

\section{Properties of the WSL wavefunction for a TB model}\label{app:WannierStarkLadder}

Here we outline some of the properties of WSL states of \eq{eq:WSL:states}.
The key element are the Bessel functions in the following representation

\begin{align*}
J_{n}(\gamma) &= \frac{1}{2\pi}\int_{-\pi}^{\pi} e^{i(k n - \gamma \sin(k))}dk\;.
\end{align*}

The mean position of the WSL state $\ket{\Psi_{m}}$ is

\begin{align*}
\avg{j}
 &= \sum_{j}    j \abs{J_{m-j}(\gamma) }^{2}\\
  &= m \underbrace{\sum_{l}     \abs{J_{l}(\gamma) }^{2}}_{\color{blue} =1}
  - \underbrace{\sum_{l}    l \abs{J_{l}(\gamma) }^{2}}_{\color{blue} =0}
= m\;.
\end{align*}

Similarly for the variance we find

\begin{align*}
\sigma^{2}&=\avg{(j-\avg{j})^{2}}\\ &= \sum_{l=-\infty}^{\infty}    l^{2} \abs{J_{l}(\gamma) }^{2}\\
&= \frac{1}{(2\pi)^{2}}\int dk dk' e^{i\gamma(\sin(k)-\sin(k'))} \sum_{l}  
 l^{2}\;e^{-il(k-k')}\\
&= - \frac{1}{2\pi}\sum_{kk'} e^{i\gamma(\sin(k)-\sin(k'))}
\big(\frac{\partial }{\partial k}\big)^{2} \delta(k-k')\;.
\end{align*}

Integration by parts eventually yields

\begin{align*}
\sigma^{2} &=\frac{\gamma^{2}}{2} = \frac{2 t^{2}}{\EE^{2}} \;.
\end{align*}

\section{Perturbation theory for an alternating potential in a Wannier-Stark model\label{app:perturbationTheory}}
Here we compute the first order energy correction of $\hat{\mathcal{H}}_{0}+
\hat{\mathcal{H}}^\text{MF}_{1}$. The terms of the Hamiltonian are defined in \eq{eq:H_TB} and \eq{eq:H_IA_MF}.

\begin{align*}
\Delta E^{(1)}_{m} &= \epsilon \sum_{l} \abs{J_{l-m}(\gamma)}^{2} e^{i\pi l}\\ 
&= e^{i\pi m} \epsilon \sum_{n} \abs{J_{n}(\gamma)}^{2} e^{i\pi n} \\
&= e^{i\pi m} \epsilon \frac{1}{(2\pi)^{2}}\sum_{kk'} e^{i\gamma(\sin(k)-\sin(k'))} \sum_{n} e^{i n (k-k'+\pi)} \\
&= \epsilon\;e^{i\pi m} \frac{1}{2\pi}\sum_{k} e^{i2\gamma \sin(k)} \\
&= \epsilon\;e^{i\pi m}  \;J_{0}(2\gamma)\;.
\end{align*}

\bibliography{1D_TB_chain_finalVersion}{}

\begin{thebibliography}{66}
\expandafter\ifx\csname natexlab\endcsname\relax\def\natexlab#1{#1}\fi
\expandafter\ifx\csname bibnamefont\endcsname\relax
  \def\bibnamefont#1{#1}\fi
\expandafter\ifx\csname bibfnamefont\endcsname\relax
  \def\bibfnamefont#1{#1}\fi
\expandafter\ifx\csname citenamefont\endcsname\relax
  \def\citenamefont#1{#1}\fi
\expandafter\ifx\csname url\endcsname\relax
  \def\url#1{\texttt{#1}}\fi
\expandafter\ifx\csname urlprefix\endcsname\relax\def\urlprefix{URL }\fi
\providecommand{\bibinfo}[2]{#2}
\providecommand{\eprint}[2][]{\url{#2}}

\bibitem[{\citenamefont{Bloch}(1928)}]{f.bl.1928}
\bibinfo{author}{\bibfnamefont{F.}~\bibnamefont{Bloch}}, \bibinfo{journal}{Z.
  Phys.} \textbf{\bibinfo{volume}{52}}, \bibinfo{pages}{555}
  (\bibinfo{year}{1928}).

\bibitem[{\citenamefont{Zener}(1934)}]{c.ze.1934}
\bibinfo{author}{\bibfnamefont{C.}~\bibnamefont{Zener}}, \bibinfo{journal}{R.
  Soc. Lond. A} \textbf{\bibinfo{volume}{145}}, \bibinfo{pages}{523}
  (\bibinfo{year}{1934}).

\bibitem[{\citenamefont{Wannier}(1960)}]{g.wa.1960}
\bibinfo{author}{\bibfnamefont{G.}~\bibnamefont{Wannier}},
  \bibinfo{journal}{Phys. Rev.} \textbf{\bibinfo{volume}{117}},
  \bibinfo{pages}{432} (\bibinfo{year}{1960}).

\bibitem[{\citenamefont{Emin and Hart}(1987)}]{emin_phonon-assisted_1987}
\bibinfo{author}{\bibfnamefont{D.}~\bibnamefont{Emin}} \bibnamefont{and}
  \bibinfo{author}{\bibfnamefont{C.~F.} \bibnamefont{Hart}},
  \bibinfo{journal}{Phys. Rev. B} \textbf{\bibinfo{volume}{36}},
  \bibinfo{pages}{2530} (\bibinfo{year}{1987}).

\bibitem[{\citenamefont{Gl{\"u}ck et~al.}(1998)\citenamefont{Gl{\"u}ck,
  Kolovsky, Korsch, and Moiseyev}}]{gluck_calculation_1998}
\bibinfo{author}{\bibfnamefont{M.}~\bibnamefont{Gl{\"u}ck}},
  \bibinfo{author}{\bibfnamefont{A.~R.} \bibnamefont{Kolovsky}},
  \bibinfo{author}{\bibfnamefont{H.~J.} \bibnamefont{Korsch}},
  \bibnamefont{and} \bibinfo{author}{\bibfnamefont{N.}~\bibnamefont{Moiseyev}},
  \bibinfo{journal}{Eur. Phys. J. D} \textbf{\bibinfo{volume}{4}},
  \bibinfo{pages}{239} (\bibinfo{year}{1998}).

\bibitem[{\citenamefont{Krieger and Iafrate}(1986)}]{kr.ia.1986}
\bibinfo{author}{\bibfnamefont{J.}~\bibnamefont{Krieger}} \bibnamefont{and}
  \bibinfo{author}{\bibfnamefont{G.}~\bibnamefont{Iafrate}},
  \bibinfo{journal}{Phys. Rev. B} \textbf{\bibinfo{volume}{33}},
  \bibinfo{pages}{5494} (\bibinfo{year}{1986}).

\bibitem[{\citenamefont{Nenciu}(1991)}]{nenciu.1991}
\bibinfo{author}{\bibfnamefont{G.}~\bibnamefont{Nenciu}},
  \bibinfo{journal}{Rev. Mod. Phys.} \textbf{\bibinfo{volume}{63}},
  \bibinfo{pages}{91} (\bibinfo{year}{1991}).

\bibitem[{\citenamefont{Bouchard and Luban}(1995)}]{bo.lu.1995}
\bibinfo{author}{\bibfnamefont{A.}~\bibnamefont{Bouchard}} \bibnamefont{and}
  \bibinfo{author}{\bibfnamefont{M.}~\bibnamefont{Luban}},
  \bibinfo{journal}{Phys. Rev. B} \textbf{\bibinfo{volume}{52}},
  \bibinfo{pages}{5105} (\bibinfo{year}{1995}).

\bibitem[{\citenamefont{Feldmann et~al.}(1992)\citenamefont{Feldmann, Leo,
  Shah, Miller, Cunningham, Meier, von Plessen, Schulze, Thomas, and
  Schmitt-Rink}}]{fe.le.1992}
\bibinfo{author}{\bibfnamefont{J.}~\bibnamefont{Feldmann}},
  \bibinfo{author}{\bibfnamefont{K.}~\bibnamefont{Leo}},
  \bibinfo{author}{\bibfnamefont{J.}~\bibnamefont{Shah}},
  \bibinfo{author}{\bibfnamefont{D.}~\bibnamefont{Miller}},
  \bibinfo{author}{\bibfnamefont{J.}~\bibnamefont{Cunningham}},
  \bibinfo{author}{\bibfnamefont{T.}~\bibnamefont{Meier}},
  \bibinfo{author}{\bibfnamefont{G.}~\bibnamefont{von Plessen}},
  \bibinfo{author}{\bibfnamefont{A.}~\bibnamefont{Schulze}},
  \bibinfo{author}{\bibfnamefont{P.}~\bibnamefont{Thomas}}, \bibnamefont{and}
  \bibinfo{author}{\bibfnamefont{S.}~\bibnamefont{Schmitt-Rink}},
  \bibinfo{journal}{Phys. Rev. B} \textbf{\bibinfo{volume}{46}},
  \bibinfo{pages}{7252} (\bibinfo{year}{1992}).

\bibitem[{\citenamefont{Waschke et~al.}(1993)\citenamefont{Waschke, Roskos,
  Schwedler, Leo, Kurz, and K{\"o}hler}}]{ch.wa.1993}
\bibinfo{author}{\bibfnamefont{C.}~\bibnamefont{Waschke}},
  \bibinfo{author}{\bibfnamefont{H.}~\bibnamefont{Roskos}},
  \bibinfo{author}{\bibfnamefont{R.}~\bibnamefont{Schwedler}},
  \bibinfo{author}{\bibfnamefont{K.}~\bibnamefont{Leo}},
  \bibinfo{author}{\bibfnamefont{H.}~\bibnamefont{Kurz}}, \bibnamefont{and}
  \bibinfo{author}{\bibfnamefont{K.}~\bibnamefont{K{\"o}hler}},
  \bibinfo{journal}{Phys. Rev. Lett.} \textbf{\bibinfo{volume}{70}},
  \bibinfo{pages}{3319} (\bibinfo{year}{1993}).

\bibitem[{\citenamefont{Yamaguchi et~al.}(1994)\citenamefont{Yamaguchi,
  Morifuji, Kubo, Taniguchi, Hamaguchi, Gmachl, and Gornik}}]{ya.mo.1994}
\bibinfo{author}{\bibfnamefont{M.}~\bibnamefont{Yamaguchi}},
  \bibinfo{author}{\bibfnamefont{M.}~\bibnamefont{Morifuji}},
  \bibinfo{author}{\bibfnamefont{H.}~\bibnamefont{Kubo}},
  \bibinfo{author}{\bibfnamefont{K.}~\bibnamefont{Taniguchi}},
  \bibinfo{author}{\bibfnamefont{C.}~\bibnamefont{Hamaguchi}},
  \bibinfo{author}{\bibfnamefont{C.}~\bibnamefont{Gmachl}}, \bibnamefont{and}
  \bibinfo{author}{\bibfnamefont{E.}~\bibnamefont{Gornik}},
  \bibinfo{journal}{Solid-State Electronics} \textbf{\bibinfo{volume}{37}},
  \bibinfo{pages}{839} (\bibinfo{year}{1994}).

\bibitem[{\citenamefont{Dahan et~al.}(1996)\citenamefont{Dahan, Peik, Reichel,
  Castin, and Salomon}}]{bd.pe.1996}
\bibinfo{author}{\bibfnamefont{M.~B.} \bibnamefont{Dahan}},
  \bibinfo{author}{\bibfnamefont{E.}~\bibnamefont{Peik}},
  \bibinfo{author}{\bibfnamefont{J.}~\bibnamefont{Reichel}},
  \bibinfo{author}{\bibfnamefont{Y.}~\bibnamefont{Castin}}, \bibnamefont{and}
  \bibinfo{author}{\bibfnamefont{C.}~\bibnamefont{Salomon}},
  \bibinfo{journal}{Phys. Rev. Lett.} \textbf{\bibinfo{volume}{76}},
  \bibinfo{pages}{4508} (\bibinfo{year}{1996}).

\bibitem[{\citenamefont{Wilkinson et~al.}(1996)\citenamefont{Wilkinson,
  Bharucha, Madison, Niu, and Raizen}}]{wi.bh.1996}
\bibinfo{author}{\bibfnamefont{S.}~\bibnamefont{Wilkinson}},
  \bibinfo{author}{\bibfnamefont{C.}~\bibnamefont{Bharucha}},
  \bibinfo{author}{\bibfnamefont{K.}~\bibnamefont{Madison}},
  \bibinfo{author}{\bibfnamefont{Q.}~\bibnamefont{Niu}}, \bibnamefont{and}
  \bibinfo{author}{\bibfnamefont{M.}~\bibnamefont{Raizen}},
  \bibinfo{journal}{Phys. Rev. Lett.} \textbf{\bibinfo{volume}{76}},
  \bibinfo{pages}{4512} (\bibinfo{year}{1996}).

\bibitem[{\citenamefont{Anderson and Kasevich}(1998)}]{an.ka.1996}
\bibinfo{author}{\bibfnamefont{B.}~\bibnamefont{Anderson}} \bibnamefont{and}
  \bibinfo{author}{\bibfnamefont{M.}~\bibnamefont{Kasevich}},
  \bibinfo{journal}{Science} \textbf{\bibinfo{volume}{282}},
  \bibinfo{pages}{1686} (\bibinfo{year}{1998}).

\bibitem[{\citenamefont{Morsch et~al.}(2001)\citenamefont{Morsch, M{\"u}ller,
  Cristiani, Ciampini, and Arimondo}}]{mo.mu.2001}
\bibinfo{author}{\bibfnamefont{O.}~\bibnamefont{Morsch}},
  \bibinfo{author}{\bibfnamefont{J.}~\bibnamefont{M{\"u}ller}},
  \bibinfo{author}{\bibfnamefont{M.}~\bibnamefont{Cristiani}},
  \bibinfo{author}{\bibfnamefont{D.}~\bibnamefont{Ciampini}}, \bibnamefont{and}
  \bibinfo{author}{\bibfnamefont{E.}~\bibnamefont{Arimondo}},
  \bibinfo{journal}{Phys. Rev. Lett.} \textbf{\bibinfo{volume}{87}},
  \bibinfo{pages}{140402} (\bibinfo{year}{2001}).

\bibitem[{\citenamefont{Hensinger et~al.}(2001)\citenamefont{Hensinger,
  H{\"a}ffner, Browaeys, Heckenberg, Helmerson, McKenzie, Milburn, Phillips,
  Rolston, Rubinsztein-Dunlop et~al.}}]{he.ha.2001}
\bibinfo{author}{\bibfnamefont{W.}~\bibnamefont{Hensinger}},
  \bibinfo{author}{\bibfnamefont{H.}~\bibnamefont{H{\"a}ffner}},
  \bibinfo{author}{\bibfnamefont{A.}~\bibnamefont{Browaeys}},
  \bibinfo{author}{\bibfnamefont{N.}~\bibnamefont{Heckenberg}},
  \bibinfo{author}{\bibfnamefont{K.}~\bibnamefont{Helmerson}},
  \bibinfo{author}{\bibfnamefont{C.}~\bibnamefont{McKenzie}},
  \bibinfo{author}{\bibfnamefont{G.}~\bibnamefont{Milburn}},
  \bibinfo{author}{\bibfnamefont{W.}~\bibnamefont{Phillips}},
  \bibinfo{author}{\bibfnamefont{S.}~\bibnamefont{Rolston}},
  \bibinfo{author}{\bibfnamefont{H.}~\bibnamefont{Rubinsztein-Dunlop}},
  \bibnamefont{et~al.}, \bibinfo{journal}{Nature}
  \textbf{\bibinfo{volume}{412}}, \bibinfo{pages}{52} (\bibinfo{year}{2001}).

\bibitem[{\citenamefont{M.Morifuji et~al.}(2002)\citenamefont{M.Morifuji, Imai,
  Hamaguchi, Carlo, Vogl, B{\"o}hm, Tr{\"a}nkle, and Weimann}}]{mo.im.2002}
\bibinfo{author}{\bibnamefont{M.Morifuji}},
  \bibinfo{author}{\bibfnamefont{T.}~\bibnamefont{Imai}},
  \bibinfo{author}{\bibfnamefont{C.}~\bibnamefont{Hamaguchi}},
  \bibinfo{author}{\bibfnamefont{A.~D.} \bibnamefont{Carlo}},
  \bibinfo{author}{\bibfnamefont{P.}~\bibnamefont{Vogl}},
  \bibinfo{author}{\bibfnamefont{G.}~\bibnamefont{B{\"o}hm}},
  \bibinfo{author}{\bibfnamefont{G.}~\bibnamefont{Tr{\"a}nkle}},
  \bibnamefont{and} \bibinfo{author}{\bibfnamefont{G.}~\bibnamefont{Weimann}},
  \bibinfo{journal}{Phys. Rev. B} \textbf{\bibinfo{volume}{65}},
  \bibinfo{pages}{233308} (\bibinfo{year}{2002}).

\bibitem[{\citenamefont{Mendez and Bastard}(1993)}]{me.ba.1993}
\bibinfo{author}{\bibfnamefont{E.}~\bibnamefont{Mendez}} \bibnamefont{and}
  \bibinfo{author}{\bibfnamefont{G.}~\bibnamefont{Bastard}},
  \bibinfo{journal}{Phys. Today} \textbf{\bibinfo{volume}{6}},
  \bibinfo{pages}{34} (\bibinfo{year}{1993}).

\bibitem[{\citenamefont{{Tarruell Leticia} et~al.}(2012)\citenamefont{{Tarruell
  Leticia}, {Greif Daniel}, {Uehlinger Thomas}, {Jotzu Gregor}, and {Esslinger
  Tilman}}}]{ta.gr.2012}
\bibinfo{author}{\bibnamefont{{Tarruell Leticia}}},
  \bibinfo{author}{\bibnamefont{{Greif Daniel}}},
  \bibinfo{author}{\bibnamefont{{Uehlinger Thomas}}},
  \bibinfo{author}{\bibnamefont{{Jotzu Gregor}}}, \bibnamefont{and}
  \bibinfo{author}{\bibnamefont{{Esslinger Tilman}}}, \bibinfo{journal}{Nature}
  \textbf{\bibinfo{volume}{483}}, \bibinfo{pages}{302} (\bibinfo{year}{2012}),
  \bibinfo{note}{10.1038/nature10871}.

\bibitem[{\citenamefont{Gl{\"u}ck et~al.}(2002)\citenamefont{Gl{\"u}ck,
  Kolovsky, and Korsch}}]{gl.ko.02}
\bibinfo{author}{\bibfnamefont{M.}~\bibnamefont{Gl{\"u}ck}},
  \bibinfo{author}{\bibfnamefont{A.}~\bibnamefont{Kolovsky}}, \bibnamefont{and}
  \bibinfo{author}{\bibfnamefont{H.}~\bibnamefont{Korsch}},
  \bibinfo{journal}{Phys. Rep.} \textbf{\bibinfo{volume}{366}},
  \bibinfo{pages}{103} (\bibinfo{year}{2002}).

\bibitem[{\citenamefont{Hartmann et~al.}(2004)\citenamefont{Hartmann, Keck,
  Korsch, and Mossmann}}]{ha.mo.04}
\bibinfo{author}{\bibfnamefont{T.}~\bibnamefont{Hartmann}},
  \bibinfo{author}{\bibfnamefont{F.}~\bibnamefont{Keck}},
  \bibinfo{author}{\bibfnamefont{H.}~\bibnamefont{Korsch}}, \bibnamefont{and}
  \bibinfo{author}{\bibfnamefont{S.}~\bibnamefont{Mossmann}},
  \bibinfo{journal}{New J. Phys.} \textbf{\bibinfo{volume}{6}},
  \bibinfo{pages}{2} (\bibinfo{year}{2004}).

\bibitem[{\citenamefont{Oka et~al.}(2003)\citenamefont{Oka, Arita, and
  Aoki}}]{ok.ar.03}
\bibinfo{author}{\bibfnamefont{T.}~\bibnamefont{Oka}},
  \bibinfo{author}{\bibfnamefont{R.}~\bibnamefont{Arita}}, \bibnamefont{and}
  \bibinfo{author}{\bibfnamefont{H.}~\bibnamefont{Aoki}},
  \bibinfo{journal}{Phys. Rev. Lett.} \textbf{\bibinfo{volume}{91}},
  \bibinfo{pages}{066406} (\bibinfo{year}{2003}).

\bibitem[{\citenamefont{Witthaut et~al.}(2005)\citenamefont{Witthaut, Werder,
  Mossmann, and H.J.Korsch}}]{wi.we.2004}
\bibinfo{author}{\bibfnamefont{D.}~\bibnamefont{Witthaut}},
  \bibinfo{author}{\bibfnamefont{M.}~\bibnamefont{Werder}},
  \bibinfo{author}{\bibfnamefont{S.}~\bibnamefont{Mossmann}}, \bibnamefont{and}
  \bibinfo{author}{\bibnamefont{H.J.Korsch}}, \bibinfo{journal}{Phys. Rev. E}
  \textbf{\bibinfo{volume}{71}}, \bibinfo{pages}{036625}
  (\bibinfo{year}{2005}).

\bibitem[{\citenamefont{Arrachea}(2004)}]{arra.2004}
\bibinfo{author}{\bibfnamefont{L.}~\bibnamefont{Arrachea}},
  \bibinfo{journal}{Phys. Rev. B} \textbf{\bibinfo{volume}{70}},
  \bibinfo{pages}{155407} (\bibinfo{year}{2004}).

\bibitem[{\citenamefont{{Di Carlo} et~al.}(1994)\citenamefont{{Di Carlo}, Vogl,
  and P{\"o}tz}}]{ca.vo.94}
\bibinfo{author}{\bibfnamefont{A.}~\bibnamefont{{Di Carlo}}},
  \bibinfo{author}{\bibfnamefont{P.}~\bibnamefont{Vogl}}, \bibnamefont{and}
  \bibinfo{author}{\bibfnamefont{W.}~\bibnamefont{P{\"o}tz}},
  \bibinfo{journal}{Phys. Rev. B} \textbf{\bibinfo{volume}{50}},
  \bibinfo{pages}{8358} (\bibinfo{year}{1994}).

\bibitem[{\citenamefont{Grecchi et~al.}(1993)\citenamefont{Grecchi, Maioli, and
  Sacchetti}}]{gr.ma.1993}
\bibinfo{author}{\bibfnamefont{V.}~\bibnamefont{Grecchi}},
  \bibinfo{author}{\bibfnamefont{M.}~\bibnamefont{Maioli}}, \bibnamefont{and}
  \bibinfo{author}{\bibfnamefont{A.}~\bibnamefont{Sacchetti}},
  \bibinfo{journal}{J. Phys. A} \textbf{\bibinfo{volume}{26}},
  \bibinfo{pages}{L379} (\bibinfo{year}{1993}).

\bibitem[{\citenamefont{Grecchi et~al.}(1994)\citenamefont{Grecchi, Maioli, and
  Sacchetti}}]{gr.ma.1994}
\bibinfo{author}{\bibfnamefont{V.}~\bibnamefont{Grecchi}},
  \bibinfo{author}{\bibfnamefont{M.}~\bibnamefont{Maioli}}, \bibnamefont{and}
  \bibinfo{author}{\bibfnamefont{A.}~\bibnamefont{Sacchetti}},
  \bibinfo{journal}{Comm. Math. Phys.} \textbf{\bibinfo{volume}{159}},
  \bibinfo{pages}{605} (\bibinfo{year}{1994}).

\bibitem[{\citenamefont{Grecchi and Sacchetti}(1995)}]{gr.sa.1995}
\bibinfo{author}{\bibfnamefont{V.}~\bibnamefont{Grecchi}} \bibnamefont{and}
  \bibinfo{author}{\bibfnamefont{A.}~\bibnamefont{Sacchetti}},
  \bibinfo{journal}{Ann. Phys.} \textbf{\bibinfo{volume}{241}},
  \bibinfo{pages}{258} (\bibinfo{year}{1995}).

\bibitem[{\citenamefont{Grecchi and
  Sacchetti}(1997{\natexlab{a}})}]{gr.sa.1997}
\bibinfo{author}{\bibfnamefont{V.}~\bibnamefont{Grecchi}} \bibnamefont{and}
  \bibinfo{author}{\bibfnamefont{A.}~\bibnamefont{Sacchetti}},
  \bibinfo{journal}{Phys. Rev. Lett.} \textbf{\bibinfo{volume}{78}},
  \bibinfo{pages}{4474} (\bibinfo{year}{1997}{\natexlab{a}}).

\bibitem[{\citenamefont{Grecchi and
  Sacchetti}(1997{\natexlab{b}})}]{gr.sa2.1997}
\bibinfo{author}{\bibfnamefont{V.}~\bibnamefont{Grecchi}} \bibnamefont{and}
  \bibinfo{author}{\bibfnamefont{A.}~\bibnamefont{Sacchetti}},
  \bibinfo{journal}{Comm. Math. Phys.} \textbf{\bibinfo{volume}{185}},
  \bibinfo{pages}{359} (\bibinfo{year}{1997}{\natexlab{b}}).

\bibitem[{\citenamefont{Grecchi and Sacchetti}(1998)}]{gr.sa.1998}
\bibinfo{author}{\bibfnamefont{V.}~\bibnamefont{Grecchi}} \bibnamefont{and}
  \bibinfo{author}{\bibfnamefont{A.}~\bibnamefont{Sacchetti}},
  \bibinfo{journal}{Comm. Math. Phys.} \textbf{\bibinfo{volume}{197}},
  \bibinfo{pages}{553} (\bibinfo{year}{1998}).

\bibitem[{\citenamefont{Buslaev and Grigis}(1998)}]{bu.gr.1998}
\bibinfo{author}{\bibfnamefont{V.}~\bibnamefont{Buslaev}} \bibnamefont{and}
  \bibinfo{author}{\bibfnamefont{A.}~\bibnamefont{Grigis}},
  \bibinfo{journal}{J. Math. Phys.} \textbf{\bibinfo{volume}{39}},
  \bibinfo{pages}{2520} (\bibinfo{year}{1998}).

\bibitem[{\citenamefont{Sacchetti}(2013)}]{sacchetti.2013}
\bibinfo{author}{\bibfnamefont{A.}~\bibnamefont{Sacchetti}}
  (\bibinfo{year}{2013}), \bibinfo{note}{arXiv:1312.6066v1}.

\bibitem[{\citenamefont{Esfahani et~al.}(2014)\citenamefont{Esfahani, Covaci,
  and Peeters}}]{es.co.2014}
\bibinfo{author}{\bibfnamefont{D.~N.} \bibnamefont{Esfahani}},
  \bibinfo{author}{\bibfnamefont{L.}~\bibnamefont{Covaci}}, \bibnamefont{and}
  \bibinfo{author}{\bibfnamefont{F.}~\bibnamefont{Peeters}},
  \bibinfo{journal}{Phys. Rev. B} \textbf{\bibinfo{volume}{90}},
  \bibinfo{pages}{205121} (\bibinfo{year}{2014}).

\bibitem[{\citenamefont{Heidrich-Meisner
  et~al.}(2010)\citenamefont{Heidrich-Meisner, Gonz\'alez, Al-Hassanieh,
  Feiguin, Rozenberg, and Dagotto}}]{he.go.10}
\bibinfo{author}{\bibfnamefont{F.}~\bibnamefont{Heidrich-Meisner}},
  \bibinfo{author}{\bibfnamefont{I.}~\bibnamefont{Gonz\'alez}},
  \bibinfo{author}{\bibfnamefont{K.~A.} \bibnamefont{Al-Hassanieh}},
  \bibinfo{author}{\bibfnamefont{A.~E.} \bibnamefont{Feiguin}},
  \bibinfo{author}{\bibfnamefont{M.~J.} \bibnamefont{Rozenberg}},
  \bibnamefont{and} \bibinfo{author}{\bibfnamefont{E.}~\bibnamefont{Dagotto}},
  \bibinfo{journal}{Phys. Rev. B} \textbf{\bibinfo{volume}{82}},
  \bibinfo{pages}{205110} (\bibinfo{year}{2010}).

\bibitem[{\citenamefont{Eckstein et~al.}(2010)\citenamefont{Eckstein, Oka, and
  Werner}}]{ec.ok.10}
\bibinfo{author}{\bibfnamefont{M.}~\bibnamefont{Eckstein}},
  \bibinfo{author}{\bibfnamefont{T.}~\bibnamefont{Oka}}, \bibnamefont{and}
  \bibinfo{author}{\bibfnamefont{P.}~\bibnamefont{Werner}},
  \bibinfo{journal}{Phys. Rev. Lett.} \textbf{\bibinfo{volume}{105}},
  \bibinfo{pages}{146404} (\bibinfo{year}{2010}).

\bibitem[{\citenamefont{Han}(2013)}]{han.13}
\bibinfo{author}{\bibfnamefont{J.~E.} \bibnamefont{Han}},
  \bibinfo{journal}{Phys. Rev. B} \textbf{\bibinfo{volume}{87}},
  \bibinfo{pages}{085119} (\bibinfo{year}{2013}).

\bibitem[{\citenamefont{Amaricci et~al.}(2012)\citenamefont{Amaricci, Weber,
  Capone, and Kotliar}}]{am.we.2012}
\bibinfo{author}{\bibfnamefont{A.}~\bibnamefont{Amaricci}},
  \bibinfo{author}{\bibfnamefont{C.}~\bibnamefont{Weber}},
  \bibinfo{author}{\bibfnamefont{M.}~\bibnamefont{Capone}}, \bibnamefont{and}
  \bibinfo{author}{\bibfnamefont{G.}~\bibnamefont{Kotliar}},
  \bibinfo{journal}{Phys. Rev. B} \textbf{\bibinfo{volume}{86}},
  \bibinfo{pages}{085110} (\bibinfo{year}{2012}).

\bibitem[{\citenamefont{Imry and Landauer}(1999)}]{im.la.1999}
\bibinfo{author}{\bibfnamefont{Y.}~\bibnamefont{Imry}} \bibnamefont{and}
  \bibinfo{author}{\bibfnamefont{R.}~\bibnamefont{Landauer}},
  \bibinfo{journal}{Rev. Mod. Phys.} \textbf{\bibinfo{volume}{71}},
  \bibinfo{pages}{S306} (\bibinfo{year}{1999}).

\bibitem[{\citenamefont{Knap et~al.}(2011)\citenamefont{Knap, von~der Linden,
  and Arrigoni}}]{knap_nonequilibrium_2011}
\bibinfo{author}{\bibfnamefont{M.}~\bibnamefont{Knap}},
  \bibinfo{author}{\bibfnamefont{W.}~\bibnamefont{von~der Linden}},
  \bibnamefont{and} \bibinfo{author}{\bibfnamefont{E.}~\bibnamefont{Arrigoni}},
  \bibinfo{journal}{Phys. Rev. B} \textbf{\bibinfo{volume}{84}},
  \bibinfo{pages}{115145} (\bibinfo{year}{2011}).

\bibitem[{\citenamefont{Nuss et~al.}(2012)\citenamefont{Nuss, Heil, Ganahl,
  Knap, Evertz, Arrigoni, and von~der Linden}}]{nuss_steady-state_2012}
\bibinfo{author}{\bibfnamefont{M.}~\bibnamefont{Nuss}},
  \bibinfo{author}{\bibfnamefont{C.}~\bibnamefont{Heil}},
  \bibinfo{author}{\bibfnamefont{M.}~\bibnamefont{Ganahl}},
  \bibinfo{author}{\bibfnamefont{M.}~\bibnamefont{Knap}},
  \bibinfo{author}{\bibfnamefont{H.~G.} \bibnamefont{Evertz}},
  \bibinfo{author}{\bibfnamefont{E.}~\bibnamefont{Arrigoni}}, \bibnamefont{and}
  \bibinfo{author}{\bibfnamefont{W.}~\bibnamefont{von~der Linden}},
  \bibinfo{journal}{Phys. Rev. B} \textbf{\bibinfo{volume}{86}},
  \bibinfo{pages}{245119} (\bibinfo{year}{2012}).

\bibitem[{\citenamefont{Aron et~al.}(2012)\citenamefont{Aron, Kotliar, and
  Weber}}]{ar.ko.12}
\bibinfo{author}{\bibfnamefont{C.}~\bibnamefont{Aron}},
  \bibinfo{author}{\bibfnamefont{G.}~\bibnamefont{Kotliar}}, \bibnamefont{and}
  \bibinfo{author}{\bibfnamefont{C.}~\bibnamefont{Weber}},
  \bibinfo{journal}{Phys. Rev. Lett.} \textbf{\bibinfo{volume}{108}},
  \bibinfo{pages}{086401} (\bibinfo{year}{2012}).

\bibitem[{\citenamefont{Han and Li}(2013)}]{ha.li.13u}
\bibinfo{author}{\bibfnamefont{J.~E.} \bibnamefont{Han}} \bibnamefont{and}
  \bibinfo{author}{\bibfnamefont{J.}~\bibnamefont{Li}} (\bibinfo{year}{2013}),
  \bibinfo{note}{arXiv:1304.4269}.

\bibitem[{\citenamefont{Tsuji et~al.}(2009)\citenamefont{Tsuji, Oka, and
  Aoki}}]{ts.ok.09}
\bibinfo{author}{\bibfnamefont{N.}~\bibnamefont{Tsuji}},
  \bibinfo{author}{\bibfnamefont{T.}~\bibnamefont{Oka}}, \bibnamefont{and}
  \bibinfo{author}{\bibfnamefont{H.}~\bibnamefont{Aoki}},
  \bibinfo{journal}{Phys. Rev. Lett.} \textbf{\bibinfo{volume}{103}},
  \bibinfo{pages}{047403} (\bibinfo{year}{2009}).

\bibitem[{\citenamefont{Roy}(2008)}]{roy.2008}
\bibinfo{author}{\bibfnamefont{D.}~\bibnamefont{Roy}}, \bibinfo{journal}{J.
  Phys.} \textbf{\bibinfo{volume}{20}}, \bibinfo{pages}{025206}
  (\bibinfo{year}{2008}).

\bibitem[{\citenamefont{Li et~al.}(2014)\citenamefont{Li, Aron, Kotliar, and
  Han}}]{kotliar.han.2014}
\bibinfo{author}{\bibfnamefont{J.}~\bibnamefont{Li}},
  \bibinfo{author}{\bibfnamefont{J.}~\bibnamefont{Aron}},
  \bibinfo{author}{\bibfnamefont{G.}~\bibnamefont{Kotliar}}, \bibnamefont{and}
  \bibinfo{author}{\bibfnamefont{J.}~\bibnamefont{Han}} (\bibinfo{year}{2014}),
  \bibinfo{note}{arXiv:1410.0626v1}.

\bibitem[{\citenamefont{Ovchinnikov and Sandalov}(1989)}]{ov.sa.89}
\bibinfo{author}{\bibfnamefont{S.~G.} \bibnamefont{Ovchinnikov}}
  \bibnamefont{and} \bibinfo{author}{\bibfnamefont{I.~S.}
  \bibnamefont{Sandalov}}, \bibinfo{journal}{Phys. C}
  \textbf{\bibinfo{volume}{161}}, \bibinfo{pages}{607} (\bibinfo{year}{1989}).

\bibitem[{\citenamefont{S{\'e}n{\'e}chal
  et~al.}(2002)\citenamefont{S{\'e}n{\'e}chal, Perez, and Plouffe}}]{se.pe.02}
\bibinfo{author}{\bibfnamefont{D.}~\bibnamefont{S{\'e}n{\'e}chal}},
  \bibinfo{author}{\bibfnamefont{D.}~\bibnamefont{Perez}}, \bibnamefont{and}
  \bibinfo{author}{\bibfnamefont{D.}~\bibnamefont{Plouffe}},
  \bibinfo{journal}{Phys. Rev. B} \textbf{\bibinfo{volume}{66}},
  \bibinfo{pages}{075129} (\bibinfo{year}{2002}).

\bibitem[{\citenamefont{Keldysh}(1965)}]{keld.65}
\bibinfo{author}{\bibfnamefont{L.~V.} \bibnamefont{Keldysh}},
  \bibinfo{journal}{Sov. Phys. JETP} \textbf{\bibinfo{volume}{20}},
  \bibinfo{pages}{1018} (\bibinfo{year}{1965}).

\bibitem[{\citenamefont{Kadanoff and Baym}(1962)}]{kad.baym}
\bibinfo{author}{\bibfnamefont{L.~P.} \bibnamefont{Kadanoff}} \bibnamefont{and}
  \bibinfo{author}{\bibfnamefont{G.}~\bibnamefont{Baym}},
  \emph{\bibinfo{title}{{Quantum Statistical Mechanics: Green's Function
  Methods in Equilibrium and Nonequilibrium Problems}}}
  (\bibinfo{publisher}{Addison-Wesley}, \bibinfo{address}{Redwood City, CA},
  \bibinfo{year}{1962}), ISBN \bibinfo{isbn}{9780201410464}.

\bibitem[{\citenamefont{Schwinger}(1961)}]{schw.61}
\bibinfo{author}{\bibfnamefont{J.}~\bibnamefont{Schwinger}},
  \bibinfo{journal}{J. Math. Phys.} \textbf{\bibinfo{volume}{2}},
  \bibinfo{pages}{407} (\bibinfo{year}{1961}).

\bibitem[{\citenamefont{Haug and Jauho}(1998)}]{ha.ja}
\bibinfo{author}{\bibfnamefont{H.}~\bibnamefont{Haug}} \bibnamefont{and}
  \bibinfo{author}{\bibfnamefont{A.-P.} \bibnamefont{Jauho}},
  \emph{\bibinfo{title}{{Quantum Kinetics in Transport and Optics of
  Semiconductors}}} (\bibinfo{publisher}{Springer},
  \bibinfo{address}{Heidelberg}, \bibinfo{year}{1998}), ISBN
  \bibinfo{isbn}{9783540735618}.

\bibitem[{\citenamefont{Rammer and Smith}(1986)}]{ra.sm.86}
\bibinfo{author}{\bibfnamefont{J.}~\bibnamefont{Rammer}} \bibnamefont{and}
  \bibinfo{author}{\bibfnamefont{H.}~\bibnamefont{Smith}},
  \bibinfo{journal}{Rev. Mod. Phys.} \textbf{\bibinfo{volume}{58}},
  \bibinfo{pages}{323} (\bibinfo{year}{1986}).

\bibitem[{\citenamefont{Economou}(2006)}]{economou.06}
\bibinfo{author}{\bibfnamefont{E.}~\bibnamefont{Economou}},
  \emph{\bibinfo{title}{{Green's Functions in Quantum Physics}}}
  (\bibinfo{publisher}{Springer}, \bibinfo{address}{Heidelberg},
  \bibinfo{year}{2006}), ISBN \bibinfo{isbn}{3540288384}.

\bibitem[{\citenamefont{Omar}(1975)}]{omar.1975}
\bibinfo{author}{\bibfnamefont{M.}~\bibnamefont{Omar}},
  \emph{\bibinfo{title}{{Elementary solid state physics: Principles and
  applications}}} (\bibinfo{publisher}{Addison-Wesley},
  \bibinfo{address}{Massachusetts}, \bibinfo{year}{1975}), ISBN
  \bibinfo{isbn}{9780201054828}.

\bibitem[{\citenamefont{Glatz et~al.}(2006)\citenamefont{Glatz, Kozub, and
  Vinokur}}]{gl.ko.2006}
\bibinfo{author}{\bibfnamefont{A.}~\bibnamefont{Glatz}},
  \bibinfo{author}{\bibfnamefont{V.}~\bibnamefont{Kozub}}, \bibnamefont{and}
  \bibinfo{author}{\bibfnamefont{V.}~\bibnamefont{Vinokur}},
  \emph{\bibinfo{title}{{Theory of Quantum Transport in Metallic and Hybrid
  Nanostructures}}} (\bibinfo{publisher}{Springer},
  \bibinfo{address}{Dordrecht}, \bibinfo{year}{2006}), ISBN
  \bibinfo{isbn}{978140204779}.

\bibitem[{\citenamefont{Kubo}(1957)}]{kubo.57}
\bibinfo{author}{\bibfnamefont{R.}~\bibnamefont{Kubo}}, \bibinfo{journal}{J.
  Phys. Soc. Jpn.} \textbf{\bibinfo{volume}{12}}, \bibinfo{pages}{570}
  (\bibinfo{year}{1957}).

\bibitem[{\citenamefont{Andronov et~al.}(2008)\citenamefont{Andronov, Dodin,
  Nozdrin, and Zinchenko}}]{andronov_transport_2008}
\bibinfo{author}{\bibfnamefont{A.~A.} \bibnamefont{Andronov}},
  \bibinfo{author}{\bibfnamefont{E.~P.} \bibnamefont{Dodin}},
  \bibinfo{author}{\bibfnamefont{Y.~N.} \bibnamefont{Nozdrin}},
  \bibnamefont{and} \bibinfo{author}{\bibfnamefont{D.~I.}
  \bibnamefont{Zinchenko}}, \bibinfo{journal}{Phys. Stat. Sol. C}
  \textbf{\bibinfo{volume}{5}}, \bibinfo{pages}{190} (\bibinfo{year}{2008}).

\bibitem[{\citenamefont{Mermin and Wagner}(1966)}]{MerminWagner}
\bibinfo{author}{\bibfnamefont{N.~D.} \bibnamefont{Mermin}} \bibnamefont{and}
  \bibinfo{author}{\bibfnamefont{H.}~\bibnamefont{Wagner}},
  \bibinfo{journal}{Phys. Rev. Lett.} \textbf{\bibinfo{volume}{17}},
  \bibinfo{pages}{1133} (\bibinfo{year}{1966}).

\bibitem[{\citenamefont{Dahnken et~al.}(2004)\citenamefont{Dahnken, Aichhorn,
  Hanke, Arrigoni, and Potthoff}}]{da.ai.04}
\bibinfo{author}{\bibfnamefont{C.}~\bibnamefont{Dahnken}},
  \bibinfo{author}{\bibfnamefont{M.}~\bibnamefont{Aichhorn}},
  \bibinfo{author}{\bibfnamefont{W.}~\bibnamefont{Hanke}},
  \bibinfo{author}{\bibfnamefont{E.}~\bibnamefont{Arrigoni}}, \bibnamefont{and}
  \bibinfo{author}{\bibfnamefont{M.}~\bibnamefont{Potthoff}},
  \bibinfo{journal}{Phys. Rev. B} \textbf{\bibinfo{volume}{70}},
  \bibinfo{pages}{245110} (\bibinfo{year}{2004}).

\bibitem[{\citenamefont{Potthoff et~al.}(2003)\citenamefont{Potthoff, Aichhorn,
  and Dahnken}}]{VCA1}
\bibinfo{author}{\bibfnamefont{M.}~\bibnamefont{Potthoff}},
  \bibinfo{author}{\bibfnamefont{M.}~\bibnamefont{Aichhorn}}, \bibnamefont{and}
  \bibinfo{author}{\bibfnamefont{C.}~\bibnamefont{Dahnken}},
  \bibinfo{journal}{Phys. Rev. Lett.} \textbf{\bibinfo{volume}{91}},
  \bibinfo{pages}{206402} (\bibinfo{year}{2003}).

\bibitem[{\citenamefont{Potthoff}(2003{\natexlab{a}})}]{VCA2}
\bibinfo{author}{\bibfnamefont{M.}~\bibnamefont{Potthoff}},
  \bibinfo{journal}{Eur. Phys. J. B} \textbf{\bibinfo{volume}{32}},
  \bibinfo{pages}{429} (\bibinfo{year}{2003}{\natexlab{a}}).

\bibitem[{\citenamefont{Potthoff}(2003{\natexlab{b}})}]{VCA3}
\bibinfo{author}{\bibfnamefont{M.}~\bibnamefont{Potthoff}},
  \bibinfo{journal}{Eur. Phys. J. B} \textbf{\bibinfo{volume}{36}},
  \bibinfo{pages}{335} (\bibinfo{year}{2003}{\natexlab{b}}).

\bibitem[{\citenamefont{Kotliar et~al.}(2001)\citenamefont{Kotliar, Savrasov,
  P\'alsson, and Biroli}}]{CDMFT}
\bibinfo{author}{\bibfnamefont{G.}~\bibnamefont{Kotliar}},
  \bibinfo{author}{\bibfnamefont{S.~Y.} \bibnamefont{Savrasov}},
  \bibinfo{author}{\bibfnamefont{G.}~\bibnamefont{P\'alsson}},
  \bibnamefont{and} \bibinfo{author}{\bibfnamefont{G.}~\bibnamefont{Biroli}},
  \bibinfo{journal}{Phys. Rev. Lett.} \textbf{\bibinfo{volume}{87}},
  \bibinfo{pages}{186401} (\bibinfo{year}{2001}).

\bibitem[{\citenamefont{Sch{\"a}fer et~al.}(2015)\citenamefont{Sch{\"a}fer,
  Geles, Rost, Rohringer, Arrigoni, Held, Bl{\"u}mer, Aichhorn, and
  Toschi}}]{sc.ge.15}
\bibinfo{author}{\bibfnamefont{T.}~\bibnamefont{Sch{\"a}fer}},
  \bibinfo{author}{\bibfnamefont{F.}~\bibnamefont{Geles}},
  \bibinfo{author}{\bibfnamefont{D.}~\bibnamefont{Rost}},
  \bibinfo{author}{\bibfnamefont{G.}~\bibnamefont{Rohringer}},
  \bibinfo{author}{\bibfnamefont{E.}~\bibnamefont{Arrigoni}},
  \bibinfo{author}{\bibfnamefont{K.}~\bibnamefont{Held}},
  \bibinfo{author}{\bibfnamefont{N.}~\bibnamefont{Bl{\"u}mer}},
  \bibinfo{author}{\bibfnamefont{M.}~\bibnamefont{Aichhorn}}, \bibnamefont{and}
  \bibinfo{author}{\bibfnamefont{A.}~\bibnamefont{Toschi}},
  \bibinfo{journal}{Phys. Rev. B} \textbf{\bibinfo{volume}{91}},
  \bibinfo{pages}{125109} (\bibinfo{year}{2015}).

\bibitem[{\citenamefont{Graf and Vogl}(1995)}]{graf_electromagnetic_1995}
\bibinfo{author}{\bibfnamefont{M.}~\bibnamefont{Graf}} \bibnamefont{and}
  \bibinfo{author}{\bibfnamefont{P.}~\bibnamefont{Vogl}},
  \bibinfo{journal}{Phys. Rev. B} \textbf{\bibinfo{volume}{51}},
  \bibinfo{pages}{4940} (\bibinfo{year}{1995}).

\end{thebibliography}

\end{document}